\documentclass{emulateapj}
\usepackage{graphicx} \usepackage{natbib} 
\shorttitle{Deep Wide-Field NIR Survey} \shortauthors{Keenan et al.}

\begin{document}

\title{An Extremely Deep, Wide-Field Near-Infrared Survey: Bright Galaxy Counts and Local Large Scale Structure}

\author{R. C. Keenan\altaffilmark{1}, L. Trouille\altaffilmark{1},
A. J. Barger\altaffilmark{1,2,3}, L. L. Cowie\altaffilmark{3},
W.-H. Wang\altaffilmark{4}}

\altaffiltext{1}{Department of Astronomy, University of Wisconsin-Madison, 475
N. Charter Street, Madison, WI 53706} 
\altaffiltext{2}{Department of Physics
and Astronomy, University of Hawaii, 2505 Correa Road, Honolulu, HI 96822}
\altaffiltext{3}{Institute for Astronomy, University of Hawaii, 2680 Woodlawn
Drive, Honolulu, HI 96822} 
\altaffiltext{4}{Institute of Astronomy and Astrophysics, Academia Sinica,
  P.O. Box 23-141, Taipei 10617, Taiwan.}

\begin{abstract} We present a deep, wide-field near-infrared (NIR) survey
over five widely separated fields at high Galactic
latitude covering a total of $\sim$ 3 deg$^2$ in  $J$, $H$, and $K_s$.  The
deepest areas of the data ($\sim$ 0.25 deg$^2$) extend to a 5 $\sigma$ limiting
magnitude of $JHK_s > 24$ in the AB magnitude system.
Although depth and area vary from field to field, the overall depth and large
area of this dataset make  it one of the deepest wide-field NIR imaging
surveys to date.  This paper discusses the observations, data reduction, and
bright galaxy counts in these fields.  We compare the slope of
the bright galaxy counts with the Two Micron All Sky Survey (2MASS) and other counts from the
literature and explore the relationship between slope and supergalactic
latitude.  The slope near the supergalactic equator is sub-Euclidean on
average pointing to the possibility of a decreasing average space density of galaxies by $\sim
10-15$\% over scales of $\sim 250-350$ Mpc.  On the contrary, the slope at high
supergalactic latitudes is strongly super-Euclidean on average suggesting an
increase in the space density of galaxies as one moves from the voids just above
and below the supergalactic plane out to distances of $\sim 250-350$
Mpc. These results suggest that local large scale structure could be
responsible for large discrepancies in the measured slope between different
studies in the past.   In addition, the local universe away from the
supergalactic plane appears to be underdense by $\sim 25-100$\% relative to the
space densities of a few hundred megaparsecs distant. 

\end{abstract} 
\maketitle

\section{Introduction}

Historically, deep and wide-area surveys in the near-infrared (NIR) have been
mutually exclusive.  Deep surveys covered only small solid angles while wide surveys
covered large solid angles, to only bright limiting
magnitudes. Both types of surveys suffer from a limited range of apparent
magnitudes over which good statistics could be obtained.  The boundaries between
the two have come closer together in
recent years with the development of large format NIR imagers capable of
covering large areas to significant depths. Bridging the gap between ``deep'' and
``wide'' NIR photometry is now becoming possible, allowing a coherent
observational picture of the history of the universe and large scale structure to be developed.   

In this paper we present a dataset composed of five large and widely separated fields at high galactic
latitudes,  including one centered on an Abell cluster (Abell 370). Here we 
present the observations, data reduction, and
bright galaxy counts in these fields.  In total, our survey consists of $\sim$ 2.75
deg$^2$ to a 5 $\sigma$ limit in a 3$\arcsec$ aperture of $JHK_{s}\sim 22-23$
and an additional $\sim 0.25$ deg$^2$ to $JHK_{s}\sim 24$,  making it
one of the deepest wide-field  surveys to date in the NIR.

The NIR samples a region of the spectral energy distribution (SED) of galaxies that
is insensitive to galaxy type over a large redshift range \citep{Cowi94}. As
such, galaxy counts in the NIR can be used as a tracer of large scale
structure without having to make assumptions about the distribution of galaxy
types in the counts.  Bright galaxy counts are expected to be
dominated by $M \leq M^* $ galaxies at low redshift ($z \leq
0.2$).  The logarithmic derivative (``slope'' hereafter) of the bright counts is expected to
follow the Euclidean prediction ($\alpha = 0.6$) until the knee of the
luminosity function induces a transition to a lower slope around an apparent
magnitude of $JHK_s \sim 17.5$ \citep{Barr09}.    
Thus, if one assumes no recent evolution in the NIR galaxy luminosity
function, then the slope of the counts curve is a measure of homogeneity in
the local universe on scales of hundreds of megaparsecs ($M^* \sim 16$ at
500 Mpc).  

Several authors have found evidence for a
large (hundreds of megaparsecs in radius)
local underdensity of galaxies using galaxy counts \citep{Huan97,
  Buss04, Frit03, Frit05}.  An underdensity on these scales
could imply that local measurements of the Hubble constant are high by some
tens of percent \citep{Huan97}. The Two Micron All Sky Survey (2MASS, \citealt{Skru06}) probes to
depths of $JHK_s\sim 15$ providing a measure of very nearby large scale
structure.  A challenge to deeper NIR surveys has been to cover a large enough
area of sky to achieve good counting statistics at magnitudes just beyond the
reach of 2MASS.   The wide-field aspect of our data allow us to
investigate the question of a possible local void by linking
our galaxy counts to those of 2MASS  and calculating
the slope of the galaxy counts curve in the local Universe.  

Catalogs for this work will be presented in an upcoming paper (R. Keenan et
al. 2010, in preparation), where we select one of the largest samples to date of Distant
Red Galaxies (DRGs, $J-K_{\rm{Vega}} < 2.3$, \citealt{Fran03}).  We will also
use this NIR survey in conjunction with deeper $K_s-$band data to constrain the
NIR cosmic infrared background.  In
addition, we are engaged in a campaign of spectroscopic followup of $H-$band
selected galaxies from this survey to constrain the local galaxy mass function
and merger rates to $z \sim 1$.  Unless otherwise noted, all magnitudes given in this paper
are in the AB magnitude system ($m_{\rm{AB}} = -2.5\log f_{\nu}-48.60$ with $f_{\nu}$ in units of
$\rm{ergs} ~\rm{cm}^{-2}~\rm{s}^{-1}~\rm{Hz}^{-1}$.

The structure of this paper is as follows:  In Section~\ref{obs} and
Section~\ref{redux} we describe the fields observed, the exposure times, the limiting
magnitudes, and the data reduction methods.
In Section~\ref{phot} we describe the photometry methods and
the completeness of the data.  We present star-galaxy separation methods and
galaxy counts in Section~\ref{numcnts}. In Section~\ref{cntscomp} we analyze
our counts with those from 2MASS and other published studies in a discussion
of galaxy counts and large scale structure.  In Section~\ref{summary} we
summarize our results. 

\begin{deluxetable*}{lccccc}[!ht]
\tabletypesize{\tiny}
\tablewidth{0pt}
\tablecaption{\label{summarytab}Coordinates, exposure times, areas, and $5~\sigma$ limiting magnitudes for each field}
\tablehead{\textbf{Field} & \textbf{CDF-N} &\textbf{A370} & \textbf{CLANS}  & \textbf{CLASXS} &\textbf{SSA13}} 
\startdata
R.A.(hh:mm:ss) & 12:36:55 & 02:39:53 & 10:46:54 & 10:34:58 & 13:12:16\\
Dec (dd:mm:ss) & 62:14:19 & -01:34:58 & 59:08:26 & 57:52:22 & 42:41:24\\ 
Galactic l (deg) &125.9 &173.0 &148.2 &151.5 &109.1\\
Galactic b (deg) &54.8  &-53.5  &51.4  &51.0  &73.8\\
Supergal. l (deg) &54.7  &302.3  &52.2  &52.3  &75.3\\
Supergal. b (deg) &11.7  &-25.7  &-1.8  &-3.9  &13.6\\
\\
\cutinhead{\textbf{$J-$band observations}}\\
Instrument & \scriptsize{CFHT WIRCAM} & \scriptsize{UH 2.2m ULBCAM} & \scriptsize{UH 2.2m ULBCAM} & \scriptsize{UH 2.2m ULBCAM} & \scriptsize{UH 2.2m ULBCAM} \\
Max. exposure time (hr)&7.7  &11.2 &4.8 &4.9 &5.7  \\
Area in square deg.&0.24  &0.55 &1.05 &0.71 &0.22    \\
3$\arcsec$ aper. 5$\sigma$ limit  &24.4 &23.5 &23.1 &23.0 &23.2  \\
\\
\cutinhead{\textbf{$H-$band observations}}\\
Instrument & \scriptsize{UH 2.2m ULBCAM} & \scriptsize{UH 2.2m ULBCAM} & \scriptsize{UH 2.2m ULBCAM} & \scriptsize{UH 2.2m ULBCAM} & \scriptsize{UH 2.2m ULBCAM} \\
Max. exposure time (hr) &24.5  & 5.9 &3.9 &2.4 &5.6 \\
Area in square deg. &0.4  &0.59 &1.24 &1.1 &0.27  \\
3$\arcsec$ aper. 5$\sigma$ limit  &24.0 &23.0 &22.5 &22.5 &22.9  \\
\\
\cutinhead{\textbf{$K_s-$band observations}}\\
Instrument & \scriptsize{CFHT WIRCAM} & \scriptsize{UKIRT WFCAM} & \scriptsize{UKIRT WFCAM} & \scriptsize{UKIRT WFCAM} & \scriptsize{UKIRT WFCAM} \\
Max. exposure time (hr) &36.8 &3.1 &15.1 &5.9 &4.0  \\
Area in square deg.&0.29 &0.87 &0.91 &1.0 &0.87    \\
3$\arcsec$ aper. 5$\sigma$ limit  &24.8 &23.0 &23.4 &23.5 &23.1  \\
\\
\enddata
\end{deluxetable*}

\section{Observations}   
\label{obs}

Our five NIR fields (CLASXS, CLANS, CDF-N, A370, and SSA13) overlap with existing data
at other wavelengths and cover a large and
diverse area of sky, which minimizes the effects of cosmic variance and large scale
structure.  We give a summary of the exposure times, field areas,
and limiting magnitudes in Table~\ref{summarytab}.  Below are descriptions of
our five fields and existing ancillary data for each.  

Our first two fields are centered on the \emph{Chandra} Large Area
Synoptic X-ray Survey (CLASXS; \citealt{Yang04,Stef04}) and the \emph{Chandra} Lockman Area North Survey
(CLANS; \citealt{Trou08,Trou09}). Each of these fields cover $\sim 1$~deg$^2$ in
$JHK_s$.  These fields  are located in the Lockman Hole region of extremely low Galactic HI column
density \citep{Lock86}.  CLASXS consists of nine overlapped $\sim$ 40~ks Chandra exposures (with
the central field observed for 73~ks) combined to create an
$\sim 0.4$~deg$^2$ image, and CLANS consists of nine separate
$\sim 70$~ks Chandra exposures combined to create an $\sim 0.6$~deg$^2$ image. \citet{Trou08,Trou09} present spectroscopic redshifts for
over half of the $>1000$ X-ray sources in these fields, and they present
photometric redshifts for almost all of the remaining sources.  They also present
optical data on these fields from CFHT MegaCam ($u,g^{\prime},i^{\prime}$ for CLASXS and
$g^{\prime},r^{\prime},i^{\prime},z^{\prime}$ for CLANS) and Subaru SuprimeCam ($B,V,R,I,z^{\prime}$ for CLASXS). These fields are
covered in the \emph{Spitzer} Wide-Area Infrared Extragalactic Survey (SWIRE,
\citealt{Lons03}) Legacy Science Program at 3.6~$\mu$m and 24~$\mu$m to limiting
fluxes of 5~$\mu$Jy and 230~$\mu$Jy, respectively.  A large portion of the CLANS
field was also recently covered in a deep 20~cm Very Large Array (VLA) survey
containing $\sim 2000$ sources to a limiting flux of $\sim 20~\mu$Jy \citep{Owen08}.    

Our third field covers a  0.25~deg$^2$ area centered on the \emph{Chandra}
Deep Field North (CDF-N, a 0.12~deg$^2$ 2~Ms \emph{Chandra} observation; \citealt{Bran01,
  Alex03}).  The entire NIR field has also
been observed in the optical at $B,V,R,I,$ and $z^{\prime}$ with Subaru
SuprimeCam and in the $U-$band with the KPNO 4~m MOSAIC instrument
\citep{Capa04}. \citet{Barg03} and  \citet{Trou08,Trou09} present
a highly complete spectroscopic redshift catalog for X-ray sources in the CDF-N along with
corresponding optical, NIR, and \emph{Spitzer} Legacy Science Program 3.6 and
24~$\mu$m photometry (M. Dickinson et al. 2009, in preparation). The CDF-N contains the Great Observatories Origins
Deep Survey North (GOODS-N; 145 arcmin$^2$ \emph{HST}
Advance Camera for Surveys observation, \citealt{Giav04}) with data at F345W,
F606W, F775W, and F850LP.  Barger et al. (2008, and references therein)
present a highly complete spectroscopic catalog of sources in the GOODS-N and
deep $K_s-$band photometry from CFHT WIRCam.  In addition, we have ultradeep
$K_s-$band image of the GOODS-N taken with the MOIRCS instrument on the Subaru
Telescope.  The GOODS-N has also been
imaged at 20~cm with the VLA by \citet{Rich00}, \citet{Bigg06}, and
G. Morrison et al. (2009, in preparation).  

Our fourth field is the Abell 370 (A370) cluster and surrounding area ($\sim
0.5$ deg$^2$).  A370 is a cluster of
richness 0 at a redshift of $z=0.37$.  We have unpublished optical data on
this field from Subaru
SuprimeCam ($B, V, R, I, z^{\prime}$). \citet{Barg01b} have studied
this field previously in optical, NIR, and radio.  We have followed up
spectroscopically thousands of sources in and around the A370 cluster, some
of which have been published for high-redshift Lyman alpha emitters
\citep{Hu02} and for low-metallicity galaxies \citep{Kaka07}.  The deep 20~cm
VLA data and more of the spectroscopic follow-up on this field will be
presented in I. Wold et al. (2010, in preparation).     

 Our fifth field is a $\sim 0.2$ deg$^2$ area 
centered on the ``Small-Survey-Area 13''(SSA13) from the Hawaii Deep
Fields described in \citet{Lill91}. \citet{Mush00} and \citet{Barg01b} have observed \emph{Chandra} X-ray
sources in this field in optical, NIR, and submillimeter wavelengths.  The
entire NIR field has also been imaged at 20~cm by \citet{Foma06}, for which
optical spectroscopy have been presented in \citet{Chap03} and \citet{Cowi04}.

\subsection{ULBCam Observations} 

We made all the $H$ and $J-$band observations (except the CDF-N $J-$band) with the Ultra Low Background Camera (ULBCam) on the UH 2.2~m
telescope. The ULBCam NIR imager was the first wide-field infrared camera. It combined four Hawaii2-RG
HgCdTe detectors to cover a $17\arcmin \times 17\arcmin$ field of view (FOV) with a plate
scale of $0\farcs25$ per pixel. ULBCam was developed as a prototype for the NIR
imager on the \emph{James Webb Space Telescope} (\emph{JWST}, \citealt{Hall04}).   

Our ULBCam observations began in September 2003.
Observations have continued since then totalling over 300 hours of observing
through February 2008.  All the data to date have been reduced and are mosaicked
into the final images. Average seeing in the final mosaics is $\sim 1\arcsec$.  

We used a $13$-step dither sequence consisting of $45\arcsec - 90\arcsec$
shifts between images having exposure times of $30-120$ s depending on the background
levels.  The maximum exposure time of 120~s kept the entire dither sequence
less than 30~min such that the coadded images from one sequence would
represent a relatively constant sky background. 

\subsection{WIRCam Observations}   

We observed the CDF-N in the $K_s-$band with the Widefield
Infrared Camera (WIRCam) at the Canada France Hawaii Telescope
3.6~m (CFHT). We observed a large area ($\sim 0.25$ deg$^2$) centered on the field.  WIRCam combines four HAWAII2-RG $2$k$ \times 2$k detectors that
cover a $20\arcmin \times 20\arcmin$ FOV with a pixel scale of $0\farcs3$.  We 
dithered images to cover the detector gaps and to provide a uniform sensitivity
distribution.  We performed most of the observations under photometric
conditions with seeing from $0\farcs6 - 1\arcsec$. The final mosaic seeing is
$\sim 0\farcs9$.  

Our group made the $K_s-$band observations in
semesters 2006A and 2007A, and we combined these data with public observations of
the GOODS-N made with the same instrument by a Canadian group led by Luc
Simard in 2006A.
Altogether the final $K_s$ image includes $\sim$ 40 hours of integration time,
making this one of the deepest wide-field $K_s$-band images ever (5 $\sigma$
limit $K_s \sim 24.8$).  The $K_s$-band catalog for this field is given in
\citet{Barg08}. 

$J-$band observations with WIRCam were obtained by a group led by Lihwai Lin in 2006A.  Our
group reduced these public data in 2008.  The final mosaic includes $\sim$ 9
hours of integration time with mosaic seeing $\sim 0\farcs8$ and a 5 $\sigma$
limiting magnitude of $J \sim 24$ .

\subsection{WFCam Observations} 

We observed four of our five fields in $K-$band
with WFCam on UKIRT (the CDF-N
is inaccessible to UKIRT because of declination constraints).  WFCam employs four HAWAII2-RG $2$k$\times 2$k
detectors spaced by 94$\%$.  This design, combined with a pixel scale of
$0\farcs4$, allows for coverage of $\sim$ 0.77 deg$^2$ in a mosaic of four pointings. Average seeing in the final WFCam mosaics is $\sim
1-1\farcs2$.

\section{Data Reduction} 
\label{redux}
Although the details of our data
reduction varied depending on the instrument used, the overall process was
essentially the same. High sky backgrounds with strong variability in the NIR
require that many relatively short exposures be coadded to create deep
images.  Dithered pointings on high Galactic latitude fields ensure that the
sky background can be effectively modelled and subtracted from mosaicked
frames.  Many short exposures also allow for the removal of cosmic rays by
determining where transient signals appear from frame to frame.  

We reduced all the ULBCam and WIRCam data using the
Interactive Data Language (IDL) SIMPLE Imaging and Mosaicking Pipeline
(SIMPLE; W.-H. Wang 2009, in preparation).  This pipeline was
designed by W.-H. Wang specifically for use in reducing dithered wide-field
near-infrared images from ground-based mosaic cameras.  In its current form,
the pipeline is optimized for blank-field extragalactic surveys where there
are no large extended objects.  Below we provide a synopsis of the pipeline
procedures, but for extensive documentation and the code itself we refer the
reader to W.-H. Wang's website\footnote{http://www.asiaa.sinica.edu.tw/$\sim$whwang/idl/SIMPLE}.  

 In the case of WFCam, we retrieved
partially reduced, stacked images from the WFCam Science Archive (WSA) and applied the final
steps of the SIMPLE pipeline to do additional background subtraction, as well
to determine the absolute astrometry and photometry for each stacked frame
before combining them into a final mosaic. 

For calibration we adopted the 2MASS zero magnitude fluxes, which are 1594 Jy, 1024 Jy and 666.7 Jy for $J$, $H$,
and $K_s$, respectively.  The conversion from  Vega to AB for 2MASS magnitudes
are $J_{\rm{AB}}= J_{\rm{Vega}}+0.894, H_{\rm{AB}}=H_{\rm{Vega}}+1.374,$ and
$K_{s,\rm{AB}}=K_{s,\rm{Vega}}+1.84$.  We analyzed all of our final images for flatness by looking at a flux ratio
between our data and 2MASS for point sources as a function of position on each image.  In this
analysis we found that our final mosaics are quite flat and uniform with no
discernable systematic differences in the flux ratio as a function of position
of the image and with typical random errors of $< 0.05$ magnitudes.  

In this paper we use the catalogs from the United Kingdom Infrared Sky Survey
(UKIDSS, \citealt{Lawr07}) for calibration checks and galaxy counts
comparison.  UKIDSS uses the UKIRT Wide Field Camera (WFCAM,
\citealt{Casa07}).  The photometric system is described in \citet{Hewe06}, and
the calibration is described in \citet{Hodg09}.  The pipeline processing and
science archive are described in M.~Irwin et al. (2009, in preparation) and
\citet{Hamb08}.  We have used data from the 3rd release, which will be
described in detail in S.~Warren et al. (2009, in preparation).  

\subsection{ULBCam Reduction Details}

We reduced images within each dither set (typically 15-30 minutes in length)
using the SIMPLE pipeline.  SIMPLE features a robust method for flat
fielding in which a sky flat is iteratively derived from dithered night sky
images.  Application of this method leads to extremely flat images after
background subtraction.  SIMPLE also corrects for image distortion in a set of
dithered images without any prior knowledge about the optics and without the
use of an astrometric catalog, which allows for accurate registration of
wide-field images. 

We applied the following procedures within the SIMPLE pipeline to generate our
reduced mosaics:  First,
we derived a median sky flat and applied it to all frames.  We then used the
SExtractor package \citep{BA96} to detect objects in each flattened frame.  We
masked detected objects to create a new median sky flat and then redid the
flatfielding.  We subtracted the sky background by
fitting a smooth polynomial surface to each masked, flattened frame.  We
removed the brightest cosmic ray hits with a 5$\times$5 pixel 
spatial sigma filter.  

Next, we again used the SExtractor package to measure object positions and
fluxes in each flattened, sky-subtracted image.  We calculated the first-order
derivative of the optical distortion function by measuring the offsets of each
object in the dither sequence as a function of location in the image.  We
obtained absolute astrometry by matching detected objects to a reference
catalog.  For this purpose we used the catalogs from the Sloan Digital Sky Survey (SDSS) \citep{York00} for all fields but Abell 370, where
USNO B1.0 was used due to the lack of SDSS coverage.  We then warped the reduced exposures to a common tangential sky plane
with subpixel accuracy.  This projection corrects for both optical distortion
and absolute astrometry.  Through this process we achieved rms astrometry
error of $\sim 0\farcs1-0\farcs2$, except in the Abell 370 field where with
the USNO B1.0 catalog we achieve an rms error of $\sim 0\farcs5$.  

We then weighted all projected images by their
relative SExtractor object fluxes to correct for variable extinction and combined them to form a dither mosaic.
In this combination we applied a sigma filter to pixels that have the same sky
position to further remove faint cosmic rays and artifacts such as reflections
inside the optics.

We applied SExtractor a third time to each dither mosaic and scaled the object
fluxes into $\mu$Jys (resulting in an AB zeropoint of 23.9 for each image) through a
comparison with 2MASS~\citep{Skru06} for point source photometry in each field
in the range $14 < JH < 16$.  This is the range over which 2MASS
reports better than $S/N=10~\sigma$ for point sources and our data are well
below saturation levels.    In the final step we coadded all reduced dither
mosaics into a large final mosaic.

\subsection{WIRCam Reduction Details}

The reduction procedure for the WIRCam data (which is only the CDF-N field)
was the same as for the ULBCam data, except for
the following details:  WIRCam shows significant crosstalk in raw frames
between the 32 readout channels.  This was removed by subtracting the median
of the 32 $64 \times 2$k channels in the object-masked image.  This process
removed the majority of crosstalk, though weak features persist around the few
brightest stars in the field.  

We obtained absolute astrometry for the WIRCam frames by matching detected
objects to a reference catalog constructed with relatively bright and compact
objects in the ACS catalog (\citealt{Giav04}; after correcting for the
$0\farcs4$ offset between ACS and radio frames) and the SuprimeCam catalog of
\citet{Capa04}. The detection limits of 2MASS in the $K_s-$band only
marginally overlap with the linear detection regime of WIRCam, and so rather than using 2MASS, we calibrated WIRCam $K_s-$band
fluxes in each individual reduced frame using our deeper Subaru MOIRCS image
as a reference \citep{Barg08}.   In the $J-$band 2MASS is deeper and we were
able to achieve a good calibration for WIRCam frames using 2MASS point sources
in the range $14<J<16$.  

Although the WIRCam and ULBCam reduction techniques were similar, the data
come from different cameras on different telescopes.  As a check for
consistency, we compared the final reduced CDF-N $J-$band WIRCam image with
the CDF-N $J-$band mosaic we generated from the ULBCam data.  The result was that for
62 relatively bright stars ($14<J<16$) we found a magnitude offset of
$J_{\rm{WIR}}-J_{\rm{ULB}}=0.003\pm 0.05$, and for 691 galaxies at $J<20$ we
found $J_{\rm{WIR}}-J_{\rm{ULB}}=0.05\pm 0.05$.  These small systematics are of
the same order of the random errors in the data, and so we consider reduced
data from these two cameras to be consistent with one another.  

\subsection{WFCam Data Reduction} 

We elected to use the stacked WFCam frames from the Cambridge Astronomical Survey Unit
(CASU) pipeline, rather than
the micro-dithered raw frames, to take advantage of the existing WFCam reduction
pipeline and the ``dribbling'' algorithm employed by
CASU to bring the relatively undersampled ($0\farcs4$ per pixel) raw frames to
higher effective resolution ($0\farcs2$ per pixel).  The following is a brief overview of the WFCam data reduction
pipeline.   Extensive documentation of the pipeline can be found at the
CASU-WFCam data processing
website\footnote{http://casu.ast.cam.ac.uk/surveys-projects/wfcam/technical/}. 

WFCam data are preprocessed in the UKIRT
summit pipeline, in which raw frames are corrected for linearity and
instrument signature,  then shipped to CASU  for further processing.  The CASU
pipeline includes astrometric alignment,
sky background subtraction, and stacking of micro-stepped images. First, dark frames are stacked to generate a master dark which is
subtracted from all frames.  Next, a flatfielding is performed using a stacked master
twilight flat.  A sky frame is then generated using a median combination with
sigma clipping of a series of dark sky science frames.  This is an iterative
process including object detection and masking to form the most robust sky
background estimation.  The resulting sky background is
then subtracted from science frames.  Individual exposures are both dithered
(tens of arcseconds offsets) and micro-stepped (subpixel offsets).  The
micro-stepping allows for the final step in the pipeline of ``dribbling''
(similar to drizzling) to stack microstepped raw frames to a finer grid. 
Processed frames are archived at the WSA.

We retrieved the stacked frames from the WSA and applied the post-stacking steps
of the SIMPLE pipeline described above to perform additional background
subtraction and spatial sigma filtering to get rid of any residuals from the
initial reduction.   We
obtained absolute astrometry by matching detected objects in each stacked frame to a reference
catalog.   We then warped the reduced stacks to a common tangential sky plane
with subpixel accuracy to correct for both optical distortion
and absolute astrometry.  Through this process, we achieved rms astrometry
error of $\sim 0\farcs1-0\farcs2$, except in the Abell 370 field where with the USNO
B1.0 catalog we achieve an rms error of $\sim 0\farcs5$.

Our CLANS field overlaps with the UKIDSS Deep Extragalactic Survey (DXS), and as such we retrieved
the public DR3 data from the WSA and added it to our final CLANS $K-$band
mosaic.  The UKIDSS DXS observations employed a very similar approach to our
observations and could be combined directly with our data using the same
reduction techniques.  

Finally, we applied SExtractor to each dither mosaic and scaled the object
fluxes into $\mu$Jys (resulting in an AB zeropoint of 23.9 for each image) through a
comparison with 2MASS~\citep{Skru06} for point source photometry in each field
in the range $14 < K_s < 16$.  In the final step we coadded all
reduced dither mosaics into a large final mosaic.  

Regarding the calibration of our WFCam data using 2MASS, the WFCam $K$ filter is a classical $K$
filter with a central wavelength of $\sim 2.2~\mu$m, while the 2MASS $K_s$
filter has a central wavelength of $\sim 2.16~\mu$m.  The difference between
photometry for stars in these two filters can be empirically quantified, as is
done in the generation of UKIDSS catalogs, where 2MASS $K_s$ is used to calibrate WFCam $K$ data via the
following relation: $K_{\rm{WFCam}} = K_{s,\rm{2MASS}} +
0.01(J_{\rm{2MASS}}-K_{s,\rm{2MASS}})$. 

We use the above UKIDSS empirical relation to calibrate our WFCam data using
2MASS.  However, for the analysis in this paper we wish to compare our WFCam
$K-$band galaxy counts with 2MASS $K_s-$band counts.  The $K_s-K$ magnitude
difference for galaxies is different than that of stars and depends both
on the shape of the galaxy's SED and its redshift.  To explore the possible
systematic and random  errors associated with generating $K_s-$band galaxy
counts from $K-$band data, we employed two methods.  

First, we used the \citet{Bruz03} template SEDs for a wide range of
stellar population ages.   We find that out to a redshift of $\sim 2.5$, the
magnitude difference $K_s-K$ has a median value of $\sim
-0.1$ with rms variability of $\pm0.05$. The difference is slightly larger ($\sim -0.15 \pm0.05$) at the
lowest redshifts.  Second, we compared magnitudes from our own photometry of
bright galaxies in our four WFCam fields (described in Section~\ref{phot}) to 2MASS
magnitudes for the same objects.  We find a $K_s-K$ difference of $\sim -0.15
\pm 0.1$ between 2MASS and our magnitudes, in good agreement with the low-redshift estimation using the \citet{Bruz03} templates described
above. Thus, in our galaxy counts and analysis, we adopt a zeropoint offset
to our $K-$band data of $-0.15$ magnitudes to bring the counts in line with the
2MASS $K_s-$band.  

In the case of our WFCam $K-$band fields, we did not have photometry from
ULBCam or WIRCam to compare with directly as a check on the different
reduction methods.  We did, however, check our photometry against the UKIDSS
catalogs in the CLANS field where we overlap and found a magnitude offset for
564 bright ($15<K<18$) stars and galaxies of
$K_{\rm{UKIDSS}}-K = 0.03\pm 0.05$ which, again, is of the same order as the
random errors in the data.  In addition, the fact that our $K-$band photometry
when compared with 2MASS $K_s$ is in agreement with the predictions from
simulations using the \citet{Bruz03} templates suggests the reductions
have performed as expected.

\section{Photometry} 
\label{phot}
\begin{deluxetable*}{lll}
\tablewidth{0pt}
\tabletypesize{\scriptsize}
\tablecaption{\label{SEtab}SExtractor Parameters}
\tablehead{Parameter & Setting\tablenotemark{a} & Comment\tablenotemark{b}}

\startdata
DETECT\_TYPE  &   CCD &            Detector type  \\
DETECT\_MINAREA  &    Variable &     Mininmum number of pixels above threshold  \\
DETECT\_THRESH &  1    &   Detection threshold ($\sigma$)  \\
ANALYSIS\_THRESH &  1  &     Limit for isophotal analysis ($\sigma$)  \\
FILTER       &   Y   &            Use filtering (Y or N)  \\
FILTER\_NAME  &   gauss\_2.0\_5x5.conv &  Filter for detection  \\
DEBLEND\_NTHRESH& 32   &    Number of deblending sub-thresholds   \\
DEBLEND\_MINCONT& 0.001 &        Minimum contrast parameter for deblending  \\
CLEAN      &   Y    &    Clean spurious detections (Y or N)  \\
CLEAN\_PARAM &    1.0   &          Cleaning efficiency  \\
MASK\_TYPE   &    CORRECT  &       Correct flux for blended objects  \\
PHOT\_APERTURES & 3, 6 &   MAG\_APER aperture diameters in arcseconds  \\
PHOT\_AUTOPARAMS& 2.5, 3.5 &       MAG\_AUTO parameters: Kron\_factor,min\_radius  \\ 
SATUR\_LEVEL  &   50000.0  &      Saturation level  \\
MAG\_ZEROPOINT &        23.9&    Magnitude zeropoint   \\
GAIN      &      1.0    &         Gain is 1 for absolute rms map  \\
PIXEL\_SCALE &    0      &        size of pixels (0=use FITS WCS info)  \\
SEEING\_FWHM  &       Variable&     Stellar FWHM in arcseconds  \\
STARNNW\_NAME &   default.nnw &    Neural-Network weight table filename  \\
BACK\_SIZE    &   48         &     Background mesh in pixels  \\
BACK\_FILTERSIZE& 8         &      Background filter  \\
BACKPHOTO\_TYPE&  LOCAL    &      Photometry background subtraction type  \\
WEIGHT\_GAIN&		N&	Gain does not vary with changes in rms noise  \\
WEIGHT\_TYPE&		MAP\_WEIGHT&	  Weighting with exposure map  \\
MEMORY\_OBJSTACK& 2000       &     number of objects in stack  \\
MEMORY\_PIXSTACK& 200000     &     number of pixels in stack  \\
MEMORY\_BUFSIZE & 1024       &     number of lines in buffer  \\

\enddata
\tablenotetext{a}{Settings listed as ``variable'' were calculated individually
  for each field in each band.  Values for the DETECT\_MINARE and SEEING\_FWHM
  are given for each field in all bands in Table \ref{area_fwhm_tab}}
\tablenotetext{b}{The SExtractor manual containing details of the SExtractor
  parameter function is available
  through TERAPIX at http://terapix.iap.fr}
\end{deluxetable*}

\begin{deluxetable*}{lcccccc}[!ht]
\tablewidth{0pt}
\tabletypesize{\tiny}
\tablecaption{\label{area_fwhm_tab}Input SEEING\_FWHM and DETECT\_MINAREA Parameters to SExtractor for Each Field in $JHK_s$}
\tablehead{\textbf{Field} & \multicolumn{2}{c}{\textbf{J}} &\multicolumn{2}{c}{\textbf{H}} & \multicolumn{2}{c}{\textbf{K}}}
\startdata
~&  SEEING\_FWHM\tablenotemark{a} & DETECT\_MINAREA\tablenotemark{a}  & SEEING\_FWHM & DETECT\_MINAREA & SEEING\_FWHM & DETECT\_MINAREA \\
CDF-N&0.88&           9&1.20&           9&0.92&           4\\
A370&1.20&          12&1.28&           9&1.08&          12\\
CLANS&1.20&          12&1.16&          16&1.12&           9\\
CLASXS&1.12&           9&1.24&          16&1.20&           9\\
SSA13&1.16&          16&1.20&          12&1.20&          16\\
\enddata
\tablenotetext{a}{Calculated as described in Section~\ref{phot}}

\end{deluxetable*}

We generated catalogs for all fields in our three NIR bands using SExtractor software version 2.4.4 \citep{BA96} for source identification and
photometry. Parameters used in the SExtractor configuration file are given in
Table \ref{SEtab}.  Parameters listed as ``variable'' in the ``setting''
column (DETECT\_MINAREA and SEEING\_FWHM) were calculated
individually for each field.  The DETECT\_MINAREA parameter sets the minimum
number of contiguous pixels required to be considered a valid source.  We
reduced DETECT\_MINAREA until a spike in the faint magnitude galaxy counts was
observed (due to spurious detections) and set the final value just above this
limit.  The SEEING\_FWHM parameter is used by SExtractor to morphologically
classify sources as pointlike or extended.  We calculated our input
SEEING\_FWHM values as described in Section~\ref{sgsep}. DETECT\_MINAREA and SEEING\_FWHM values for each field are listed in Table
\ref{area_fwhm_tab}.   MAG\_ZEROPOINT values were set to 23.9 in all
cases corresponding to units of $\mu$Jy in the images.  

We masked areas near bright objects to exclude areas of possible saturation or
diffraction spikes.  We define bright objects to have a USNO-B1.0 $R-$band
magnitude $< 14.5$.  Following the
method used in \citet{Capa04}, we mask an area around such objects within a radius defined by (where $R$ stands for
$R-$band magnitude): 
\begin{equation} 
Radius=291.2-34.55\times R+1.06\times R^2
\end{equation}
We also masked regions of low exposure time in the mosaic.  We defined low
exposure time to be any area having less than 50\% of the median exposure
time in the mosaic.  

As a measure of the depth of our fields, we calculated 5 $\sigma$ limiting magnitudes (listed
in Table~\ref{summarytab}) by measuring the rms noise in randomly scattered 3$\arcsec$ apertures at
blank locations in the image.  We did this by scattering $10^4$ apertures
randomly and retrieving rms noise values for those that fell at least
3$\arcsec$ away from detected objects.  We retrieved the  median rms value in
all scattered apertures and recorded this as the noise background of the image
(i.e., 1~$\sigma$).

\vspace{5mm}

\section{Galaxy Counts} 
\label{numcnts}
\begin{figure*}
\begin{center}
\includegraphics[width=150mm]{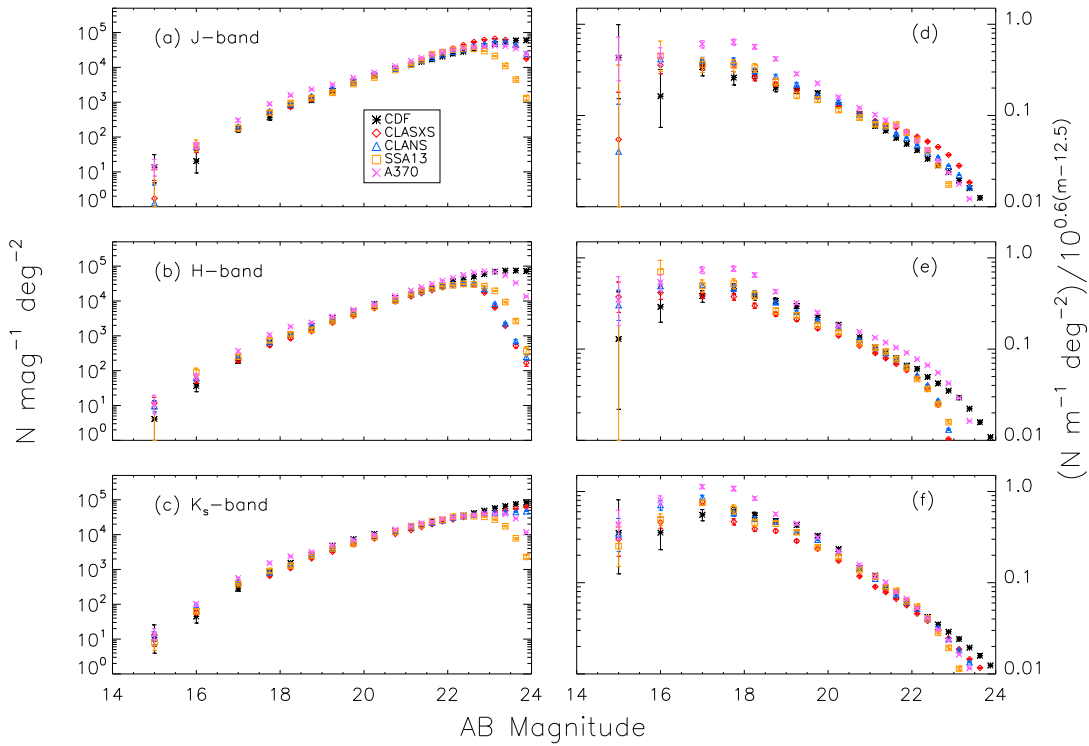}
\caption{\label{mycounts}(a-c)   Raw galaxy counts  in each of our five
  fields.  $J-$band counts are shown in (a), $H-$band in (b) and $K_s-$band in (c).  Error bars represent
  1 $\sigma$ Poisson fluctuations. (d-f) The same data as in (a-c) divided
  through by an arbitrary normalized Euclidean model to expand the ordinate.}
\end{center}
\end{figure*}

Our survey covers $\sim 3$ deg$^2$ in $J$ and $H$ and nearly $4$ deg$^2$ in
$K_s$.  However, only 1.8 deg$^2$ is covered simultaneously in all 3 bands.  As
shown in Section~\ref{sgsep}, we reliably separate stars from galaxies at
$JHK_s<17.5$ by morphology alone.  As such, for $JHK_s<17.5$ we count galaxies
over the entire extent of our survey to achieve the best counting statistics
possible.  For $JHK_s>17.5$ we switch to counting galaxies over only the 1.8
deg$^2$ of overlap in all three bands in order to also employ color-selection
techniques in separating stars from galaxies.

We constructed the galaxy counts by first making several cuts to the
output catalog of SExtractor.  We excluded objects in masked areas as
described in Section~\ref{phot}.  We identified stars by the methods described
in Section~\ref{sgsep}.   The raw counts for $JHK_s < 24$ are shown in
Figure~\ref{mycounts}, and the raw counts for $JHK_s < 22$, where all fields
are highly complete, are given in tabular form in Tables~\ref{jcounts}-\ref{kcounts}.

We binned
identified galaxies in variable sized bins using large (1 magnitude) bins at
bright magnitudes and decreasing to small (0.25 magnitude) bins at faint
magnitudes.  This scheme of decreasing the bin size toward fainter magnitudes allowed us to
provide better statistics on the bright end of the galaxy counts curve and
better resolution on the faint end.  We divided the bin totals by the unmasked area of the image.
We determined Poissonian galaxy count errors (1 $\sigma$) using the
counting confidence limits of \citet{Gehr86}.     

\begin{deluxetable*}{cccccc}
\tabletypesize{\tiny}
\tablewidth{0pt}
\tablecaption{\label{jcounts}$J-$band Raw Galaxy Counts}
\tablehead{ Mag(AB) & \multicolumn{5}{c}{log$(N)~mag^{-1} \rm{deg}^{-2} \pm 1\sigma$\tablenotemark{a}}} 
\startdata
~&CDF-N&CLANS&CLASXS&SSA13&A370\\
15.000&1.134$\pm$ 0.361&0.105$\pm$ 0.518&0.236$\pm$ 0.518&0$\pm$ 0.265&1.133$\pm$ 0.225\\
16.000&1.310$\pm$ 0.293&1.717$\pm$ 0.063&1.651$\pm$ 0.077&1.751$\pm$ 0.163&1.756$\pm$ 0.085\\
17.000&2.230$\pm$ 0.079&2.303$\pm$ 0.033&2.241$\pm$ 0.041&2.260$\pm$ 0.073&2.483$\pm$ 0.039\\
17.750&2.564$\pm$ 0.067&2.744$\pm$ 0.035&2.716$\pm$ 0.035&2.698$\pm$ 0.060&2.954$\pm$ 0.031\\
18.250&2.907$\pm$ 0.046&2.939$\pm$ 0.028&2.862$\pm$ 0.030&2.974$\pm$ 0.045&3.203$\pm$ 0.023\\
18.750&3.054$\pm$ 0.039&3.178$\pm$ 0.022&3.092$\pm$ 0.023&3.119$\pm$ 0.038&3.371$\pm$ 0.019\\
19.250&3.321$\pm$ 0.029&3.391$\pm$ 0.017&3.339$\pm$ 0.017&3.270$\pm$ 0.032&3.506$\pm$ 0.016\\
19.750&3.593$\pm$ 0.021&3.568$\pm$ 0.014&3.550$\pm$ 0.014&3.522$\pm$ 0.024&3.701$\pm$ 0.013\\
20.250&3.763$\pm$ 0.018&3.800$\pm$ 0.010&3.750$\pm$ 0.011&3.708$\pm$ 0.020&3.851$\pm$ 0.011\\
20.750&3.962$\pm$ 0.014&3.962$\pm$ 0.009&3.971$\pm$ 0.008&3.925$\pm$ 0.015&4.033$\pm$ 0.009\\
21.125&4.060$\pm$ 0.018&4.109$\pm$ 0.010&4.115$\pm$ 0.010&4.077$\pm$ 0.018&4.183$\pm$ 0.010\\
21.375&4.159$\pm$ 0.016&4.200$\pm$ 0.009&4.218$\pm$ 0.009&4.208$\pm$ 0.015&4.273$\pm$ 0.009\\
21.625&4.228$\pm$ 0.015&4.277$\pm$ 0.008&4.344$\pm$ 0.008&4.374$\pm$ 0.013&4.363$\pm$ 0.008\\
21.875&4.311$\pm$ 0.013&4.379$\pm$ 0.007&4.448$\pm$ 0.007&4.436$\pm$ 0.012&4.444$\pm$ 0.008\\
\enddata
\tablenotetext{a}{Errors represent Poissonian confidence limits of \citet{Gehr86}}
\end{deluxetable*}

\begin{deluxetable*}{cccccc}
\tabletypesize{\tiny}
\tablewidth{0pt}
\tablecaption{\label{hcounts}$H-$band Raw Galaxy Counts}
\tablehead{ Mag(AB) & \multicolumn{5}{c}{log$(N)~mag^{-1} \rm{deg}^{-2} \pm 1\sigma$\tablenotemark{a}}} 
\startdata
~&CDF-N&CLANS&CLASXS&SSA13&A370\\
15.000&0.609$\pm$ 0.518&0.983$\pm$ 0.163&1.071$\pm$ 0.163&0$\pm$ 0.265&1.038$\pm$ 0.255\\
16.000&1.563$\pm$ 0.163&1.792$\pm$ 0.053&1.719$\pm$ 0.063&1.948$\pm$ 0.124&1.834$\pm$ 0.079\\
17.000&2.281$\pm$ 0.059&2.399$\pm$ 0.027&2.290$\pm$ 0.034&2.395$\pm$ 0.062&2.566$\pm$ 0.035\\
17.750&2.813$\pm$ 0.051&2.818$\pm$ 0.033&2.725$\pm$ 0.035&2.868$\pm$ 0.050&3.031$\pm$ 0.028\\
18.250&3.034$\pm$ 0.040&3.065$\pm$ 0.025&2.925$\pm$ 0.028&3.042$\pm$ 0.041&3.261$\pm$ 0.022\\
18.750&3.278$\pm$ 0.031&3.262$\pm$ 0.020&3.133$\pm$ 0.022&3.169$\pm$ 0.036&3.380$\pm$ 0.019\\
19.250&3.520$\pm$ 0.023&3.458$\pm$ 0.016&3.375$\pm$ 0.017&3.419$\pm$ 0.027&3.555$\pm$ 0.015\\
19.750&3.696$\pm$ 0.019&3.665$\pm$ 0.012&3.576$\pm$ 0.013&3.613$\pm$ 0.022&3.750$\pm$ 0.012\\
20.250&3.917$\pm$ 0.015&3.882$\pm$ 0.009&3.797$\pm$ 0.010&3.835$\pm$ 0.017&3.907$\pm$ 0.010\\
20.750&4.080$\pm$ 0.012&4.051$\pm$ 0.008&3.986$\pm$ 0.008&4.011$\pm$ 0.014&4.138$\pm$ 0.008\\
21.125&4.193$\pm$ 0.015&4.187$\pm$ 0.009&4.131$\pm$ 0.010&4.191$\pm$ 0.016&4.300$\pm$ 0.009\\
21.375&4.289$\pm$ 0.014&4.281$\pm$ 0.008&4.224$\pm$ 0.009&4.298$\pm$ 0.014&4.397$\pm$ 0.008\\
21.625&4.375$\pm$ 0.012&4.366$\pm$ 0.008&4.312$\pm$ 0.008&4.365$\pm$ 0.013&4.492$\pm$ 0.007\\
21.875&4.444$\pm$ 0.011&4.443$\pm$ 0.007&4.394$\pm$ 0.007&4.419$\pm$ 0.012&4.585$\pm$ 0.006\\
\enddata
\tablenotetext{a}{Errors represent Poissonian confidence limits of \citet{Gehr86}}
\end{deluxetable*}

\begin{deluxetable*}{cccccc}
\tabletypesize{\tiny}
\tablewidth{0pt}
\tablecaption{\label{kcounts}$K-$band Raw Galaxy Counts}
\tablehead{ Mag(AB) & \multicolumn{5}{c}{log$(N)~mag^{-1} \rm{deg}^{-2} \pm 1\sigma$\tablenotemark{a}}} 
\startdata
~&CDF-N&CLANS&CLASXS&SSA13&A370\\
15.000&1.047$\pm$ 0.361&1.028$\pm$ 0.176&0.976$\pm$ 0.176&0.900$\pm$ 0.204&1.139$\pm$ 0.155\\
16.000&1.649$\pm$ 0.176&1.951$\pm$ 0.050&1.763$\pm$ 0.057&1.794$\pm$ 0.059&2.003$\pm$ 0.048\\
17.000&2.444$\pm$ 0.057&2.634$\pm$ 0.023&2.585$\pm$ 0.023&2.575$\pm$ 0.025&2.752$\pm$ 0.020\\
17.750&2.934$\pm$ 0.045&2.914$\pm$ 0.029&2.818$\pm$ 0.032&2.955$\pm$ 0.046&3.182$\pm$ 0.024\\
18.250&3.188$\pm$ 0.034&3.152$\pm$ 0.022&3.038$\pm$ 0.025&3.107$\pm$ 0.039&3.375$\pm$ 0.019\\
18.750&3.425$\pm$ 0.026&3.418$\pm$ 0.016&3.317$\pm$ 0.018&3.396$\pm$ 0.028&3.502$\pm$ 0.016\\
19.250&3.683$\pm$ 0.019&3.609$\pm$ 0.013&3.507$\pm$ 0.014&3.602$\pm$ 0.022&3.698$\pm$ 0.013\\
19.750&3.864$\pm$ 0.016&3.821$\pm$ 0.010&3.722$\pm$ 0.011&3.736$\pm$ 0.019&3.846$\pm$ 0.011\\
20.250&4.018$\pm$ 0.013&3.978$\pm$ 0.008&3.890$\pm$ 0.009&3.932$\pm$ 0.015&3.995$\pm$ 0.009\\
20.750&4.110$\pm$ 0.012&4.108$\pm$ 0.007&4.021$\pm$ 0.008&4.087$\pm$ 0.013&4.147$\pm$ 0.008\\
21.125&4.252$\pm$ 0.014&4.219$\pm$ 0.009&4.130$\pm$ 0.010&4.239$\pm$ 0.015&4.243$\pm$ 0.010\\
21.375&4.271$\pm$ 0.014&4.288$\pm$ 0.008&4.222$\pm$ 0.009&4.276$\pm$ 0.014&4.329$\pm$ 0.009\\
21.625&4.370$\pm$ 0.012&4.344$\pm$ 0.008&4.298$\pm$ 0.008&4.385$\pm$ 0.013&4.372$\pm$ 0.008\\
21.875&4.423$\pm$ 0.012&4.412$\pm$ 0.007&4.378$\pm$ 0.007&4.420$\pm$ 0.012&4.449$\pm$ 0.008\\
\enddata
\tablenotetext{a}{Errors represent Poissonian confidence limits of \citet{Gehr86}}
\end{deluxetable*}

We then performed a weighted average of galaxy counts over all fields. We
created the weighted average by taking into account
Poisson errors (weight $=1/\sigma^2$).  This ensures that results from the
widest fields in our survey carry more weight on the bright end.    

\subsection{Star-Galaxy Separation}
\label{sgsep}

\begin{figure}
\begin{center}
\includegraphics[width=60mm]{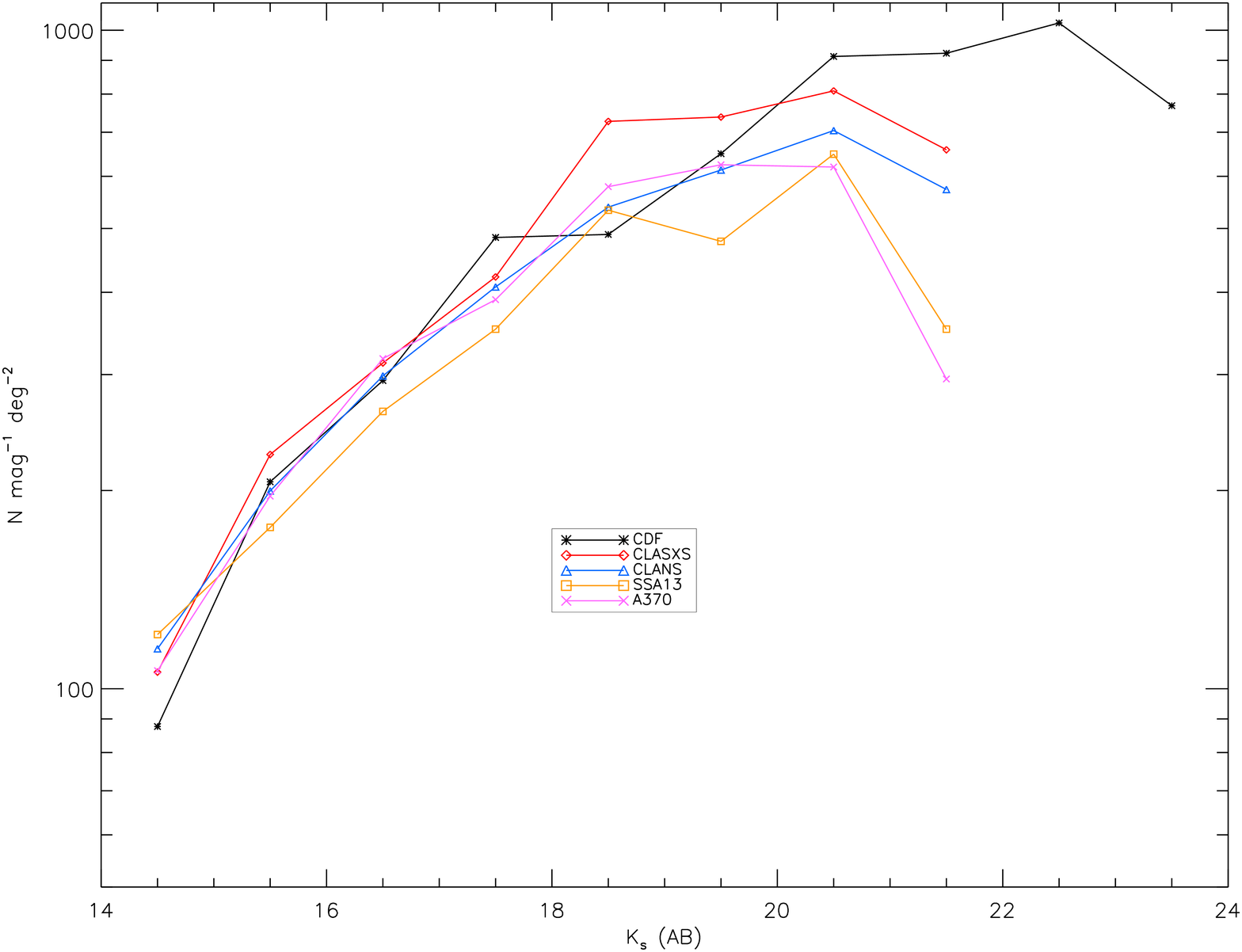}
\caption{\label{starcount}Star counts for each of our fields using the
  combination of morphology and color criteria
described in Section~\ref{sgsep}}
\end{center}
\end{figure}

We used a combination of morphology and color to separate stars from galaxies
for our galaxy counts.  Our methods include use of the
SExtractor ``CLASS\_STAR'' parameter, the ratio of semimajor to semiminor axes
of  objects, and $J-K_s$ color to separate stars from galaxies.  We honed our
methods on a group of spectroscopically identified stars and galaxies in the
GOODS-N (a subfield of the CDF-N). This involved the use of $J$ and $K_s-$band
catalogs from this study and the GOODS-N redshift catalog from
\citet{Barg08}.  Finally, as an independent test of our methods, in the four
of our five fields (all but A370) where we overlap with the SDSS, we compared
our classified stars having $JHK_s < 19$ with the morphological
classifications for these objects in the SDSS catalogs.  The results of these
methods are discussed below.  Our final star counts
generated using these methods are shown in Figure~\ref{starcount}

The CLASS\_STAR parameter is an output of SExtractor to describe
whether a given object looks like a point source or an extended source.  The
CLASS\_STAR output contains a value between 0 (extended object) and 1 (point source) and is assigned by
SExtractor to each object detected in the image.  CLASS\_STAR can be used to separate stars
from galaxies at bright magnitudes, but the accuracy of this
parameter breaks down at fainter magnitudes.  Where that breakdown occurs
depends on the depth of the field and the SEEING\_FWHM  input parameter (see
Table \ref{SEtab}).

\begin{figure}
\begin{center}
\includegraphics[width=80mm]{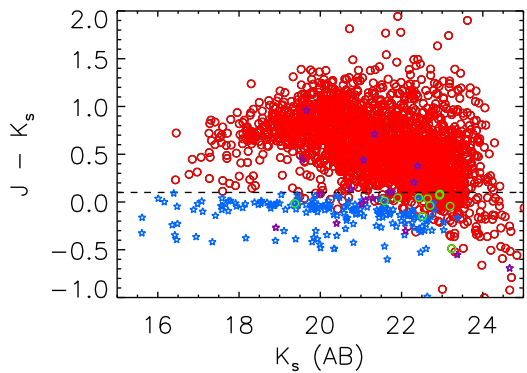}
\caption{\label{JK}$J-K_s$ color-magnitude plot for spectroscopically identified stars and galaxies in the
GOODS-N region (a subfield of the CDF-N) from the catalog of \citet{Barg08}.  Accurately identified stars are shown
in blue.  Stars we fail to identify are shown in purple.  Accurately
identified galaxies are shown in red circles and galaxies misidentified as
stars are in green circles.  The vast majority of stars
are found below the line $J-K_s=0.1$.}
\end{center}
\end{figure} 

\begin{figure}
\begin{center}
\includegraphics[width=80mm]{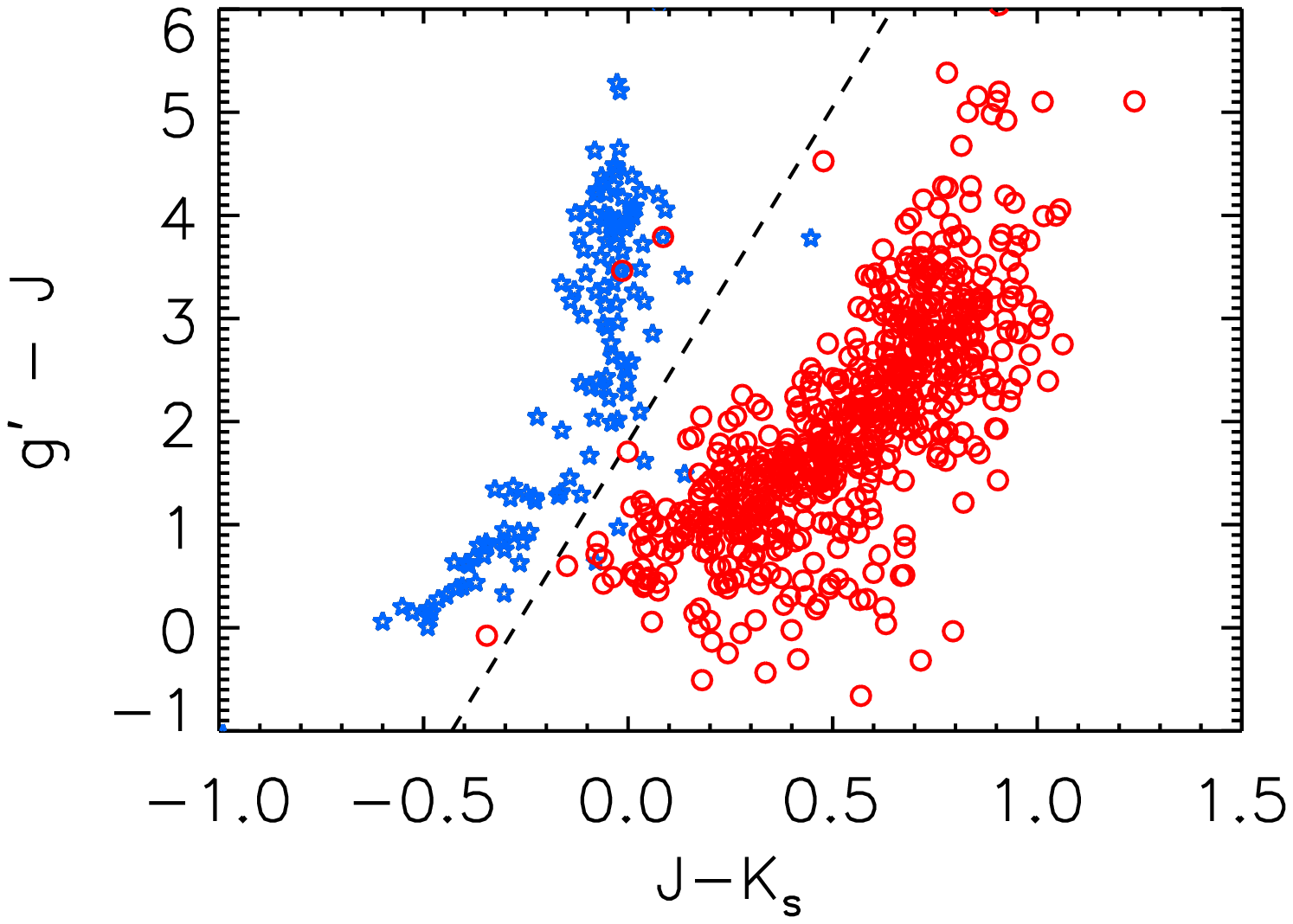}
\caption{\label{GJK}A color-color plot demonstrating the utility of a
  combination of the SDSS $g^{\prime}$ filter with the $J$ and $K_s$ filters in the
  NIR for star-galaxy separation.  Stars are shown as blue stars and galaxies
  as red circles from the spectroscopic catalog of \citet{Barg08}.  We use the
$g^{\prime}$$JK_s$ color-color data from our fields that overlap with the SDSS to
investigate discrepancies between our star-galaxy separation and that of the SDSS, as described in~Section~\ref{sgsep}}
\end{center}
\end{figure} 

The SEEING\_FWHM input parameter is provided to SExtractor by the user
to describe the average point source Full-Width-Half-Maximum (FWHM) in the
image.   This parameter is critical for
SExtractor to accurately identify stars with its output parameter
CLASS\_STAR.   

We determined the optimal value for the SEEING\_FWHM input parameter by first
running SExtractor on an image to determine the FWHM of point sources in the
image (using an initial guess at the image seeing for the input SEEING\_FWHM).
We then took the median FWHM for selected point sources (CLASS\_STAR $>0.95$) as the final input
value for the SEEING\_FWHM parameter.  Values calculated in this way for the
SEEING\_FWHM parameter in each field are listed in Table \ref{area_fwhm_tab}.

We found that on occasion for bright galaxies the CLASS\_STAR parameter
misidentified galaxies as stars that were clearly galaxies by eye (2 out of
125 galaxies at $K_s < 19$ in the GOODS-N catalog).  We resolved this issue by
constraining the ratio of semimajor to semiminor axis for stars ($a/b \leq
1.3$).   

Using the star
identification criteria of CLASS\_STAR $> 0.7$ for $JK_s<17.5$ and
CLASS\_STAR $>0.9$  and $a/b \leq 1.3$ for  $17.5<JK_s<19$, we correctly identified
$100\%$ of stars for $JK_s<19$ from the spectroscopic sample of \citet{Barg08}.  

 To extend our star-galaxy separation to fainter magnitudes, we used a
combination of the SExtractor CLASS\_STAR parameter and the $J-K_s$ color of
objects.  We determined that a color cut of $J-K_s < 0.1$ and CLASS\_STAR
$>0.95$ for $19<JK_s<24$ accurately identified $>90$\% of stars while only
misclassifying $<1$\% of galaxies as stars.  Because galaxies dominate the
counts at these faint magnitudes, the combination of these effects introduces
less than 1\% error in the counts in the range $19<JK_s<24$.  These results are
displayed in Figure~\ref{JK} where $K_s-$band magnitude is plotted against $J-K_s$
color for the \citet{Barg08} sample.  Blue stars show correctly identified
stars, red circles show correctly identified galaxies, and purple stars and
green circles show misclassified stars and galaxies, respectively.  

In our fields that overlap with SDSS (all but A370), we were able to
perform a check on our bright $(JHK_s<19)$ star selection by cross-correlating
with objects classified morphologically as point sources in SDSS.  We
looked at two things in this analysis.  First, we searched for objects that we
identifed as stars which were classified as extended objects in SDSS.  Second,
we searched for objects that we found to be galaxies that were classified as point sources in SDSS.  

For $JHK_s<17.5$,
we found that none of our galaxies were listed as point
sources in SDSS.  However, a handful (5-15) of our stars in each field were
listed as extended sources in SDSS.  We
first investigated this discrepancy by eye and found that many of these
objects were clearly point sources in our images identifiable by their
diffraction spikes.  This is perhaps due to the fact that our images are
much deeper than SDSS, or these are late-type stars that are brighter in
the NIR than in the optical.  Other sources appeared to be close
binary point sources in our images, perhaps unresolved by the SDSS.  Given
that the majority of discrepant point sources appear to be misclassifications
in the SDSS, we conclude that we have done a robust job of star-galaxy
separation for $JHK_s < 17.5$.  

In the range $17.5 < JHK_s < 19$, we found $>90$\% agreement between objects we
classfied as stars and point sources
identified by SDSS, as well as for objects we classified as galaxies and
their SDSS counterparts.  To investigate discrepant sources, we employed
$g^{\prime}$(sdss) $-J$ versus $J-K_s$ color-color diagrams.  The utility of such diagrams for
separating stars and galaxies is shown in Figure~\ref{GJK} using the
\citet{Barg08} sample as an example, where SDSS counterparts are found for 131
stars and 647 galaxies.  Using this sample, we determined that a color cut of
$g^{\prime} - J > 6.5\times(J-K_s)+1.8$ was appropriate for selecting the vast
majority of stars.  

We plotted objects we identified as stars, but whose SDSS counterparts were
classified as extended on the $g^{\prime} JK_s$ plane and found that the majority matched
the color criterion for stars defined above.  Those that did not meet the
criterion typically had very high SExtractor CLASS\_STAR parameters in our
catalogs and, as such, are likely candidates to be late-type M and L dwarf
stars. 

Finally, we plotted objects that we identified as galaxies but whose SDSS
counterparts were classified as point sources on the $g^{\prime} JK_s$ plane.  We
found a minority of these objects did, in fact, meet the $g^{\prime} JK_s$ color
criterion for stars. In this case, we flagged these objects as stars in our
catalogs. Stars identified in this way through comparison with the SDSS were
very few in number and only affect the galaxy counts in the $17.5<JHK_s<19$
range on the $\sim 0-3$\% level.  

The SDSS does not cover our A370 field, so we do not have the opportunity to
compare our star-galaxy separation with an independent optical morphology
classification like we have done for the other fields. Based on our comparison
with the SDSS in other fields, we expect that stars we may have missed in the
A370 field could contribute to an excess on the few percent level to counts in
the $17.5<JHK_s<19.$ range.   
 
As seen in Figure~\ref{JK}, there are stars that will be missed by our color
cut in $J-K_s$.  These will tend to be M and L dwarf stars because
they are redder than the rest of the main sequence population (T dwarfs, in
fact, have $J-K_s < 0$ like typical main sequence stars).  However, we would
expect to identify the majority of M and L dwarfs as stars based on morphology
alone for $K_s<19$ (as demonstrated above),
corresponding to distances of out to $\sim 100$ parsecs. Beyond this we would
miss these faint dwarf stars in our star removal.  

The space density of M and L dwarfs is poorly constrained at large
heliocentric distances, but \citet{Stan08} and \citet{Pirz09} have shown that M and L dwarfs may exist in
large numbers out to several kpc into the Galactic halo.  The star counts for
M dwarfs from these two groups roughly agree and appear flat out to very faint
magnitudes at levels of $\sim 100-200$ mag$^{-1}$ deg$^2$.  

In our galaxy
counts, assuming the worst case scenario, this could represent a 15\%
systematic error in the brightest bin ($K_s = 19-20$), for which we require $J-K_s <
0.1$.  Assuming flat star counts for these late-type dwarf stars, this error
would drop to $<10$\% at $K_s=20.5$ and $<5$\% at $K_s=21.5$.  The focus of this
paper is the bright galaxy counts for which we believe our star-galaxy
separation is robust, but we alert the reader that the counts remain
uncorrected for late-type dwarf stars fainter than $JHK_s=19$.

\section{Bright Galaxy Counts and Local Large Scale Structure}
\label{cntscomp}
\begin{figure}
\begin{center}
\includegraphics[width=80mm]{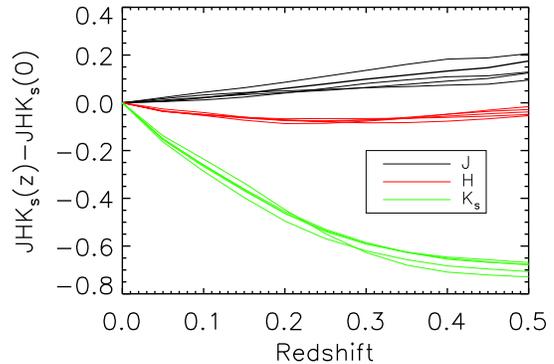}
\caption{\label{kcorrections} NIR $K-$corrections are shown as a function of redshift
  in $J$ (black), $H$ (red), and $K_s$ (green), from \citet{Mann01}.  Five galaxy types (E, S0, Sa, Sb, Sc)
  are shown (though not identified individually) to demonstrate the similarity
  in each NIR bandpass
 of the $K-$correction for different galaxy types.  Because of this similarity
 the uncertainty in type mixture in our galaxy counts only leads to
 uncertainties of $<0.005$ in the determination of the expected Euclidean slope.}
\end{center}
\end{figure} 

\begin{figure}
\begin{center}
\includegraphics[width=80mm]{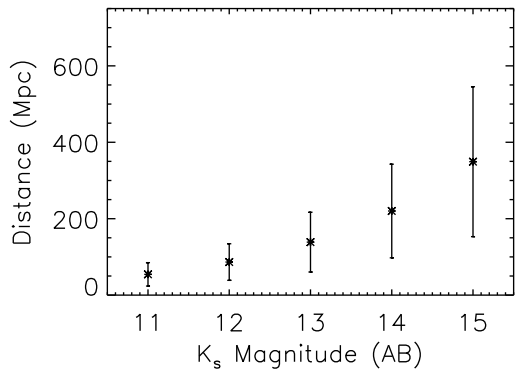}
\caption{\label{kvdist} The distance of galaxies at a given
  $K_s-$band magnitude (the distributions are very similar in $H$ and $J-$band) assuming homogeneity and a non-evolving NIR luminosity
  function.  This distance distribution
  was calculated using the NIR luminosity functions of \citet{Jone09}
  and the $K-$corrections of \citet{Mann01}.  The plotted points show the peak
  of the distance distribution ($M^*$) in a given magnitude bin, and the error bars
  show one standard deviation above and below the peak in each bin.   }
\end{center}
\end{figure}

In the past two decades, numerous NIR
galaxy counts studies have been published.  Typically new data are overplotted
with a selection of previous works to demonstrate how the various studies
agree or disagree in a given bandpass.  At bright magnitudes, factors of two
or more discrepancy between any two studies are not uncommon in these
comparisons.  A lack of knowledge about the proper normalization of the galaxy
counts curve makes it difficult to determine when such differences are
actually due to variations in large scale structure and not other effects, such
as differences or inconsistencies in photometry.  

The slope of the bright galaxy counts curve, on the other hand, is a measure
of local large scale structure that is independent of the normalization.
 In a static, homogeneous, Euclidean universe, galaxy counts would follow the curve
$N=A\times 10^{\alpha m},$ where $A$ is a normalization constant, $\alpha =
0.6$, and $m$ is apparent magnitude.  A slope of $\alpha = 0.6$ is
also expected for the bright end of the
galaxy counts curve in an expanding, homogeneous universe
where  $K-$corrections are small and nearly independent of galaxy type and
where cosmological effects on the luminosity distance are small.

\citet{Barr09} show that for galaxy counts derived from \citet{Sche76} luminosity
functions, the bright counts slope is expected to follow the Euclidean
prediction and be dominated by $M \leq M^*$ galaxies until a transition to
lower slope is induced by the knee in the luminosity function around $JHK_s \sim
17.5$. 

The bright counts slope is then modified by the $K-$corrections in a
given bandpass, which act effectively as a changing $M^*$.  Optical
$K-$corrections are large and galaxy type dependent, and thus 
considerations of large scale structure based on the slope of optical galaxy
counts are plagued with uncertainties about the exact mix of galaxy types in
the counts.  NIR $K-$corrections are smaller than in the optical and nearly
type independent.  Figure~\ref{kcorrections} shows the low-redshift $K-$corrections from
 \citet{Mann01} ($J$ in black, $H$ in red, and $K_s$ in green). These are shown
 for five galaxy types (E, S0, Sa, Sb, Sc), demonstrating the near type
 independence of the NIR $K-$corrections.

It is straightforward to determine the effect of the
$K-$corrections of \citet{Mann01} on the theoretical slope of the galaxy counts curve in a
homogeneous distribution of galaxies in the nearby expanding universe ($z <
0.1$, $cz = H_0D$).  Assuming $H_0 =
70~\rm{km}~\rm{s}^{-1}~\rm{Mpc}^{-1}$ and an equal mix of all five galaxy
types shown in Figure~\ref{kcorrections}, the slope for the $J-$band should be $\alpha
\sim 0.57$, for the $H-$band $\alpha
\sim 0.58$, and for the $K_s-$band, $\alpha \sim 0.61$.  
Because of the similarity of the $K-$corrections across galaxy types the uncertainty introduced by the lack of knowledge of the exact type
mixture in the calculation of the expected slope is less than $\pm 0.005$.  This result is also independent of the shape of
the galaxy luminosity function if one assumes no galaxy evolution at
$z<0.1$. In Figure~\ref{kvdist} we show the distribution in distance of
galaxies for a range of $K_s-$band magnitudes.  

In what follows, we consider departures of $\sim 0.01-0.1$ from the
Euclidean prediction .  If we ignore evolution and treat the slope as a direct
measure of local large scale structure, these departures can be thought of as changes in
the space density of galaxies as one moves further out into the local
universe.  For example, a difference of $0.01$ from the Euclidean predication
measured over an interval of three magnitudes would roughly imply a galaxy space density
change of $\sim 7$\% over the range of distances sampled at those
magnitudes. As shown in Figure~\ref{kvdist}, the range of distances sampled in
any given magnitude bin increases toward fainter magnitude, adding uncertainty
to the physical significance of a change in slope, but if the luminosity
function is truly a constant over the range measured, then the slope is
a measure of the space density of galaxies over that range.

\citet{Madd90} found a steeper than
Euclidean slope in $B-$band counts over the 4300 deg$^2$ of the Automated
Plate Measuring (APM) galaxy
survey and suggested rapid galaxy evolution at low redshifts as an
explanation.  \citet{Koo92} point out that dramatic collective evolution near
the present epoch would be unusual, and they instead attribute the
steep slope of the APM survey counts to some combination of model
uncertainties and normal evolution.  \citet{Shan90} point out that the steep
galaxy counts slope could be due to clustering if we are located inside a
large ($\sim 150h^{-1}$ Mpc) region of $\sim 50$\% underdensity.  However, as
mentioned above, optical $K-$corretions are highly type dependent, so a
measurement of the slope in the NIR is required for a more reliable test of
local large scale structure.

\citet{Huan97} found a steeper than Euclidean slope
($\alpha = 0.69$) in
$K_s-$band counts over a collection of fields totalling $\sim 10$ deg$^2$, but
they point out that the amount of evolution needed to account for the effect, particularly in the NIR at low
redshifts, seems unlikely.  The alternative to rapid evolution cited by
\citet{Huan97}, is a $\sim 50$\% local underdensity of
galaxies stretching 300$h^{-1}$~Mpc in extent.  They point out that this would
mean local measurements of the cosmological mass density, $\Omega_0$, would be
low by the same factor and that local measurements of $H_0$ could be 
high by as much as 33\%.  Evidence for such a ``local void'' using 2MASS, the
Two-degree Field Galaxy Redshift Survey (2dFGRS), and other data, along with
galaxy counts models, has been presented in \citet{Buss04} and
\citet{Frit03,Frit05}.  In what follows, we look at the bright galaxy counts
slope of our data in combination with 2MASS, the variability in the slope as a
function of position on the sky, and possible implications for local large
scale structure.

\subsection{Galaxy Counts in the 2MASS XSC}
\begin{figure*}
\begin{center}
\includegraphics[width=150mm]{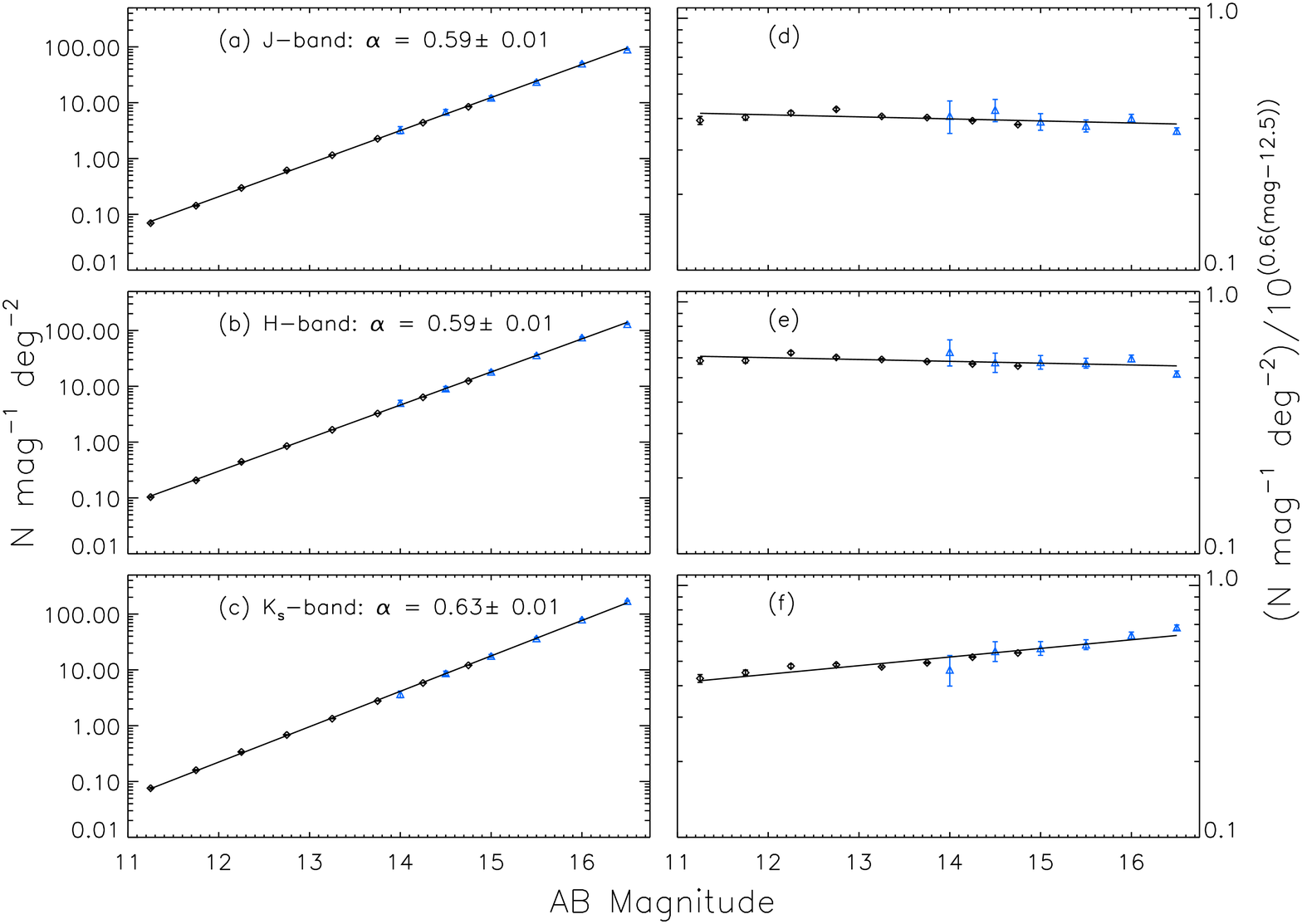}
\caption{\label{masscnts}(a-c) Average counts for 2MASS $|\rm{b}|>30$ (black
  diamonds) and for 
  2MASS-6x Lockman Hole (blue triangles). 
  Error bars show 1 $\sigma$ Poisson fluctuations (smaller than plot symbols in
  a-c). (d-f) are the same data as (a-c) after having been divided through by a
  normalized Euclidean model (i.e., $[N~mag^{-1} \rm{deg}^{-2}]/10^{0.6(\rm{m}-12.5)}$). Solid lines show an error-weighted least-squares
fit to the counts.  The calculated slopes  are listed in the plots. }
\end{center}
\end{figure*}

\begin{figure*}
\begin{center}
\includegraphics[width=150mm]{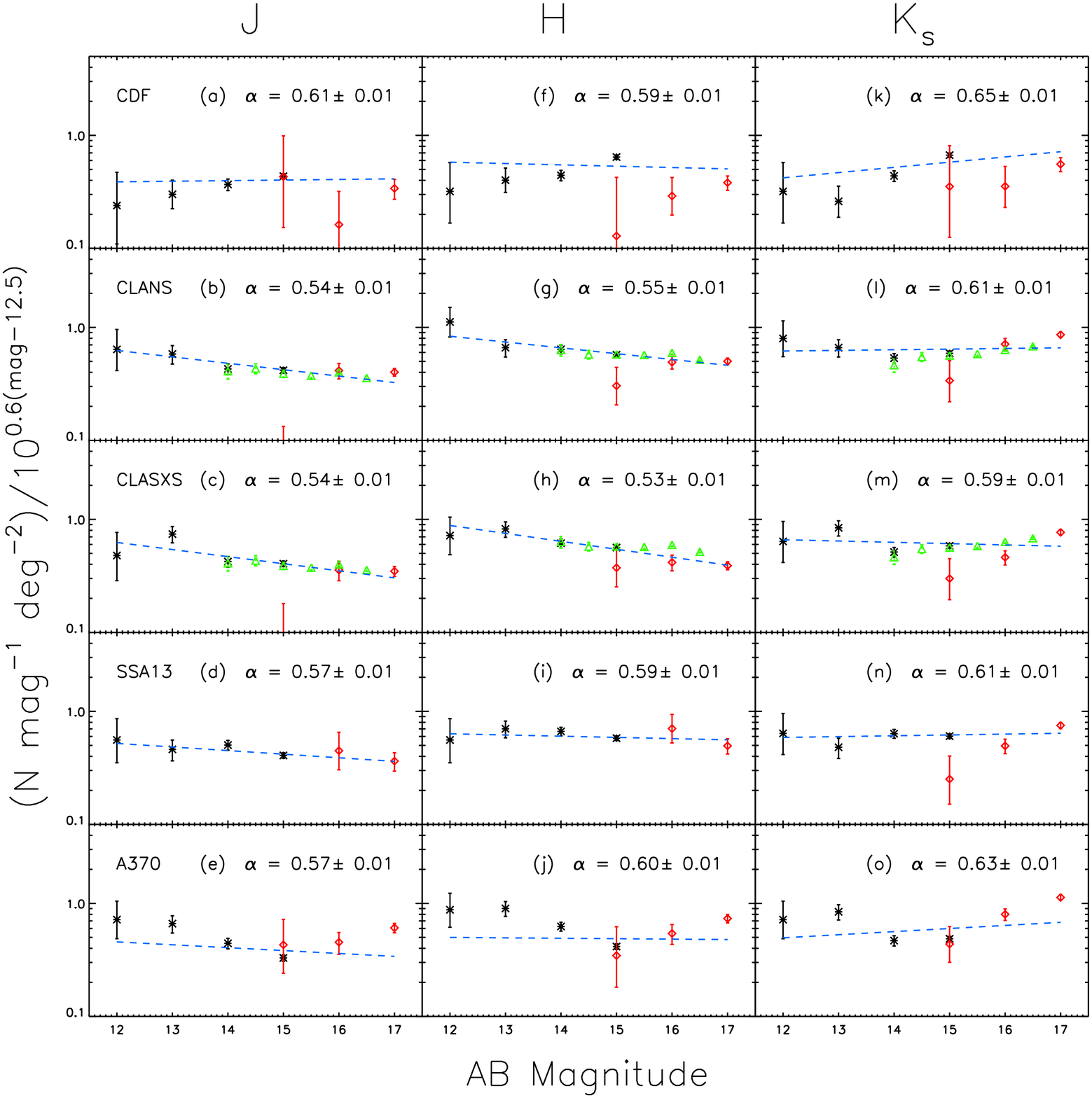}
\caption{\label{masscomp} (a-e) $J-$band counts for 2MASS (black asterisks)
  vs. counts from this study in each field (red diamonds).  Green triangles
  represent the counts from the 2MASS-6x survey, which overlaps with the CLANS
  and CLASXS fields.  (f-j) $H-$band counts in the same format.  (k-o)
  $K_s-$band counts in the same format.  The blue dashed lines represent an
  error-weighted least squares fit to the data, and the calculated slope is
listed in each plot.  The counts have been divided through by a normalized
Euclidean model to expand the ordinate.}
\end{center}
\end{figure*}

  We downloaded the entire 2MASS extended source catalog (XSC) and
2MASS-6x XSC from the NASA/IPAC Infrared Science
Archive\footnote{http://irsa.ipac.caltech.edu}. We constructed 2MASS galaxy counts from the catalog over an
area of $\sim 20,000$ deg$^2$ representing the entire 2MASS XSC for Galactic
latitude $|b|>$ 30 (2$\pi$ Steradians).  We limited this 2MASS sample to
$JHK_s<15.5$ to stay within the all sky catalog's completeness
limits.  We excluded known 2MASS Galactic extended sources from the
counts using the Galactic extended source catalog available at NASA/IPAC website.   As shown in Figure~\ref{kvdist}, galaxy counts from 2MASS from 11-15$^{th}$
magnitude are sampling $M^*$ galaxies in the local universe at distances of
$\sim 50-350$ Mpc.

We also constructed galaxy counts from the 2MASS-6x \citep{Beic03}
catalog of extended sources in the $\sim 25$ deg$^2$ centered on the
Lockman Hole area of low galactic HI column density \citep{Lock86}.  These data represent 25 deg$^2$ where 2MASS
observed six times as long as the all sky catalog exposure times, thus making
the 2MASS-6x catalog roughly one magnitude deeper in $J, H$ and $K_s$.  We were
able to further ``deepen'' the 2MASS-6x Lockman Hole catalog because our deep
data in the CLANS and CLASXS fields overlaps 100\% with the 2MASS-6x Lockman
Hole.  As such, we were able to measure completeness in the
2MASS-6x catalog, and correct the counts to $JHK_s=16.75$.  Thus, we use the 2MASS and 2MASS-6x
catalogs to firmly establish average galaxy counts on the bright end. The
average 2MASS galaxy counts for $|b|>30$~deg are shown in Figure~\ref{masscnts}.

The solid lines in Figure~\ref{masscnts} show an error-weighted least
squares fit to the data.  In the left panels the counts are shown in the
standard $N$ vs. magnitude form, and in the right panels the counts are shown divided through by
a normalized Euclidean model.  For this global average of galaxy counts in $J$,
$H$, and $K_s$ we find $\alpha = 0.59, 0.59$, and $0.63~(\sim
\pm 0.01$ in all
three bands), respectively.  These values are all 0.01-0.02 above the Euclidean
prediction, representing a possible galaxy space density increase toward fainter
magnitudes of 7-15\% over the magnitude range sampled.  
 
Next, we extracted 2MASS counts for areas of $\sim 25$ deg$^2$ centered on
each of our fields.  In Figure~\ref{masscomp} we show our
raw counts (red diamonds) for $JHK_s<17.5$ plotted with 2MASS counts (black asterisks) from each of these 25
deg$^2$ subfields.  The dashed lines show an error-weighted least-squares
fit to the data.  In the CLANS and CLASXS fields we include our counts
extracted from the 2MASS-6x survey (green triangles), because these two fields lie within the
2MASS-6x Lockman Hole.  The counts have been divided through by a normalized
Euclidean model ($\alpha = 0.6$).  The calculated slopes are tabulated in the plots.  We note
that in many cases fitting to only the data points from this study would
yield a much steeper slope than that obtained with the inclusion of 2MASS.  In
this measurement of the local slope at our field positions, we find a mix of
values ranging from sub-Euclidean to super-Euclidean.  We note that the most
sub-Euclidean slopes are found in our fields near the supergalactic equator
(CLANS and CLASXS).  We return to this point in Section~\ref{supgalcnts}.

\begin{figure}
\begin{center}
\includegraphics[width=80mm]{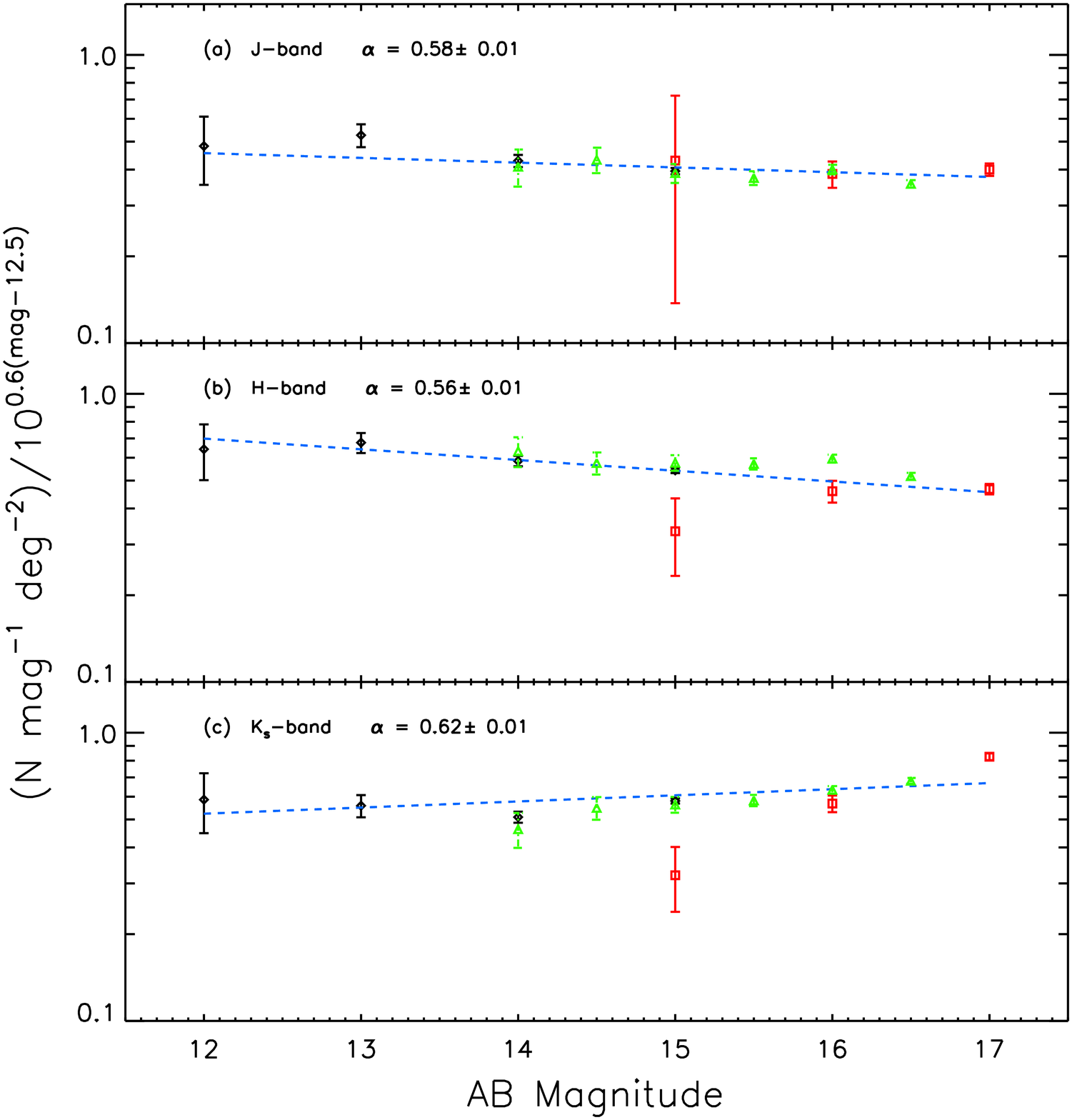}
\caption{\label{localavg} Average $J$, $H$, and $K_s-$band counts for 2MASS
  subfields of 25~deg$^2$ each centered on our fields (black
  diamonds), 2MASS-6x Lockman Hole (green triangles), and average counts for
  this study (red squares).  The counts have been divided through by a
  normalized Euclidean model to expand the ordinate.}
\end{center}
\end{figure}

In Figure~\ref{localavg}, we show the error-weighted average of our counts
(red squares) alongside the error-weighted average of the 2MASS counts (black diamonds) extracted from the 25
deg$^2$ subfields centered on our fields (for a total of 125 deg$^2$), as well
as the counts from the 2MASS-6x Lockman Hole (green triangles).  Again, we fit the slope with an
error-weighted least-squares method and the slopes are listed in the
plots.  We find that, on average, when combined with 2MASS, our counts yield a
Euclidean slope within our error bars of $\pm 0.01$.  We note that taken on their own, our
counts in both the
$H$ and $K_s-$bands would appear super-Euclidean.  This is likely due
simply to a lack of bright objects in the fields ($HK_s\sim 15$), leading to
large random errors in the brightest bin.   

\subsection{Galaxy Counts and the Supergalactic Plane}
\label{supgalcnts}

\begin{figure}
\begin{center}
\includegraphics[width=80mm]{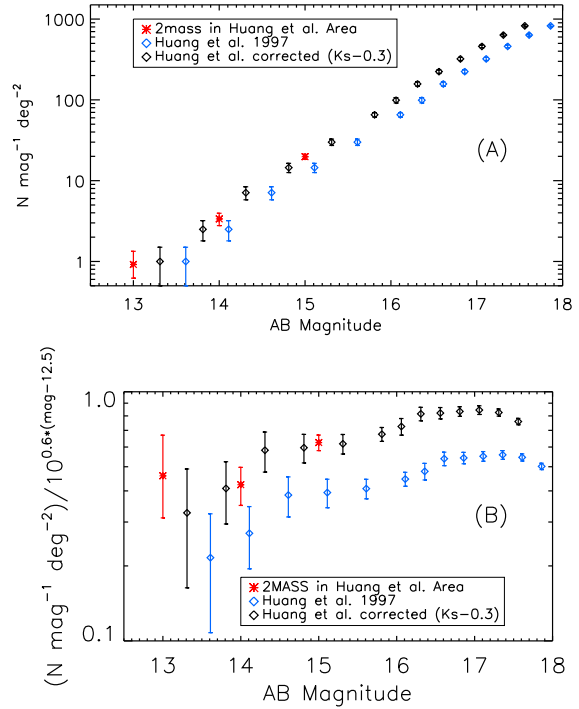}
\caption{\label{huang_corr}(a) The \citet{Huan97} $K_s$ counts are shown as blue
diamonds.  The 2MASS counts were constructed for the \citet{Huan97} fields for
comparison and are shown as red asterisks.  After applying a zeropoint
correction ($K_s=K_s-0.3$; see text) the Huang et al. counts (black diamonds)
align well with the 2MASS
$K_s$ counts from the same area.  We adopt this adjustment as a correction to
the Huang et al. counts, bringing them into good agreement with other studies over the
same magnitude range.  Error bars show 1 $\sigma$ Poisson fluctuations. (b) The
same data as in (a) but divided through by a normalized
Euclidean model (i.e. $[\rm{N} mag^{-1} deg^{-2}]/10^{0.6(K_s-12.5)}$).   }
\end{center}
\end{figure} 

\begin{figure}
\begin{center}
\includegraphics[width=80mm]{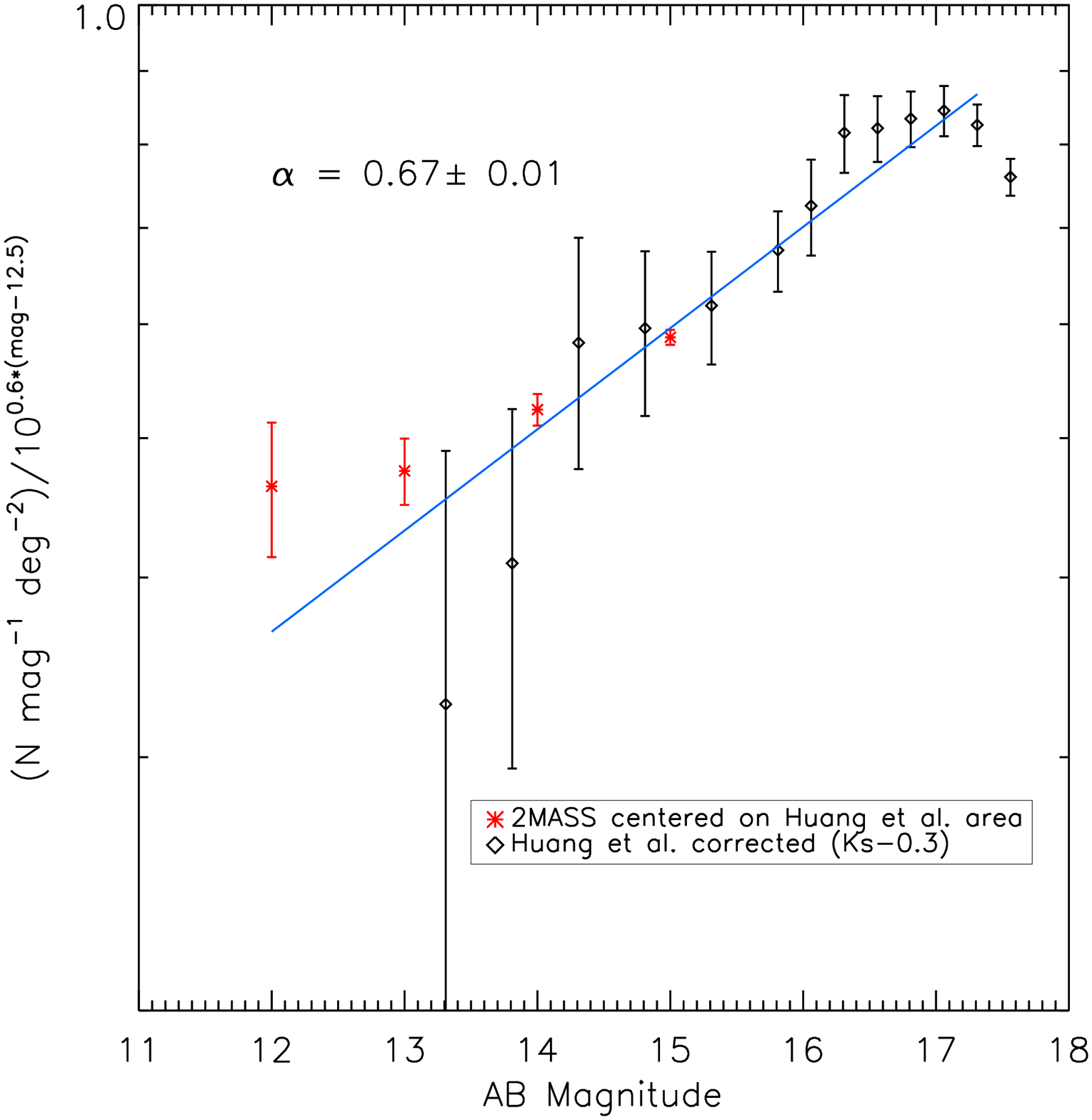}
\caption{\label{huang_v_mass}(a) The \citet{Huan97} $K_s$ counts are shown as black
diamonds after a zeropoint correction ($K_s-0.3$; see text).  The 2MASS counts
were constructed from 25 deg$^2$ subfields (375 deg$^2$ total) centered on the
\citet{Huan97} fields for comparison and are shown as red asterisks.  The
line shows an error-weighted least-squares fit to the data.  We
find $\alpha = 0.67$ in this fit, a slightly lower number than was found
for the \citet{Huan97} data alone ($\alpha=0.69$), but still distinctly
super-Euclidean.  } 
\end{center}
\end{figure}

\begin{figure}
\begin{center}
\includegraphics[width=80mm]{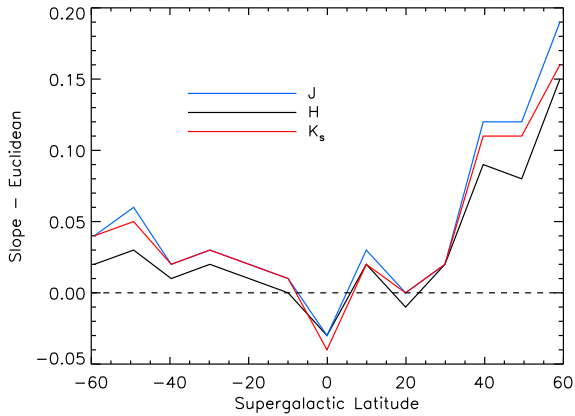}
\caption{\label{supgalslope} Average slope as a function of supergalactic
  latitude (SGB) for the 2MASS counts ($|\rm{b}|>30; 10.5<JHK_s<15.5$).  The
  expected Euclidean slope has been subtracted from the measured values such
  that the dashed horizontal line at 0 represents the Euclidean expectation in
  all bands. $J-$band appears in blue, $H$ in black, and $K_s$ in red.  Northern SGB
  slopes show a strong trend toward super-Euclidean, generally consistent with
  results from \citet{Huan97} and \citet{Gard93}.  The only distinctly
  sub-Euclidean slopes appear at SGB $\sim 0$, consistent with our findings in
the CLANS and CLASXS fields which lie near the supergalactic equator. }  
\end{center}
\end{figure}

\begin{figure}
\begin{center}
\includegraphics[width=80mm]{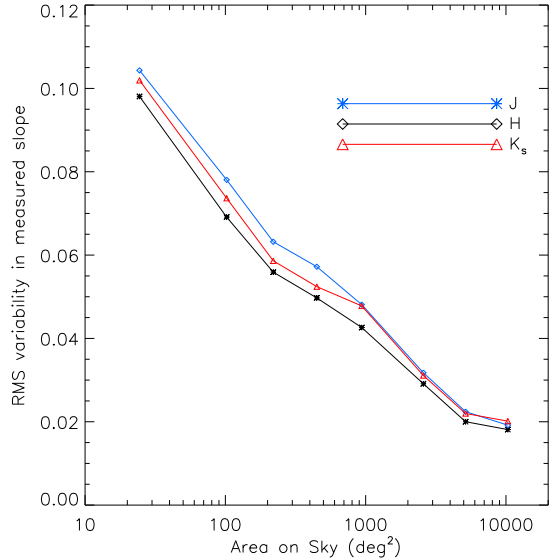}
\caption{\label{areaverr} 
The 2MASS $|\rm{b}|>30$ sky was divided up into
  increasingly larger patches of area on the sky ranging from $\sim 25-10000~\rm{deg}^2$ to test for when the rms variability in slope between sub-areas
  was reduced to the order of the variability induced by the supergalactic
  plane alone.  The rms variability in measured slope is shown as a function
  of the area of the patches of sky.   The result is that for patches of sky
  greater than $\sim 1000$ deg$^2$, the variability in the slope can be
  accounted for by the supergalactic plane alone, such that other local large
  scale structure has been sufficiently averaged out. }
\end{center}
\end{figure}

\begin{figure*}
\begin{center}
\includegraphics[width=150mm]{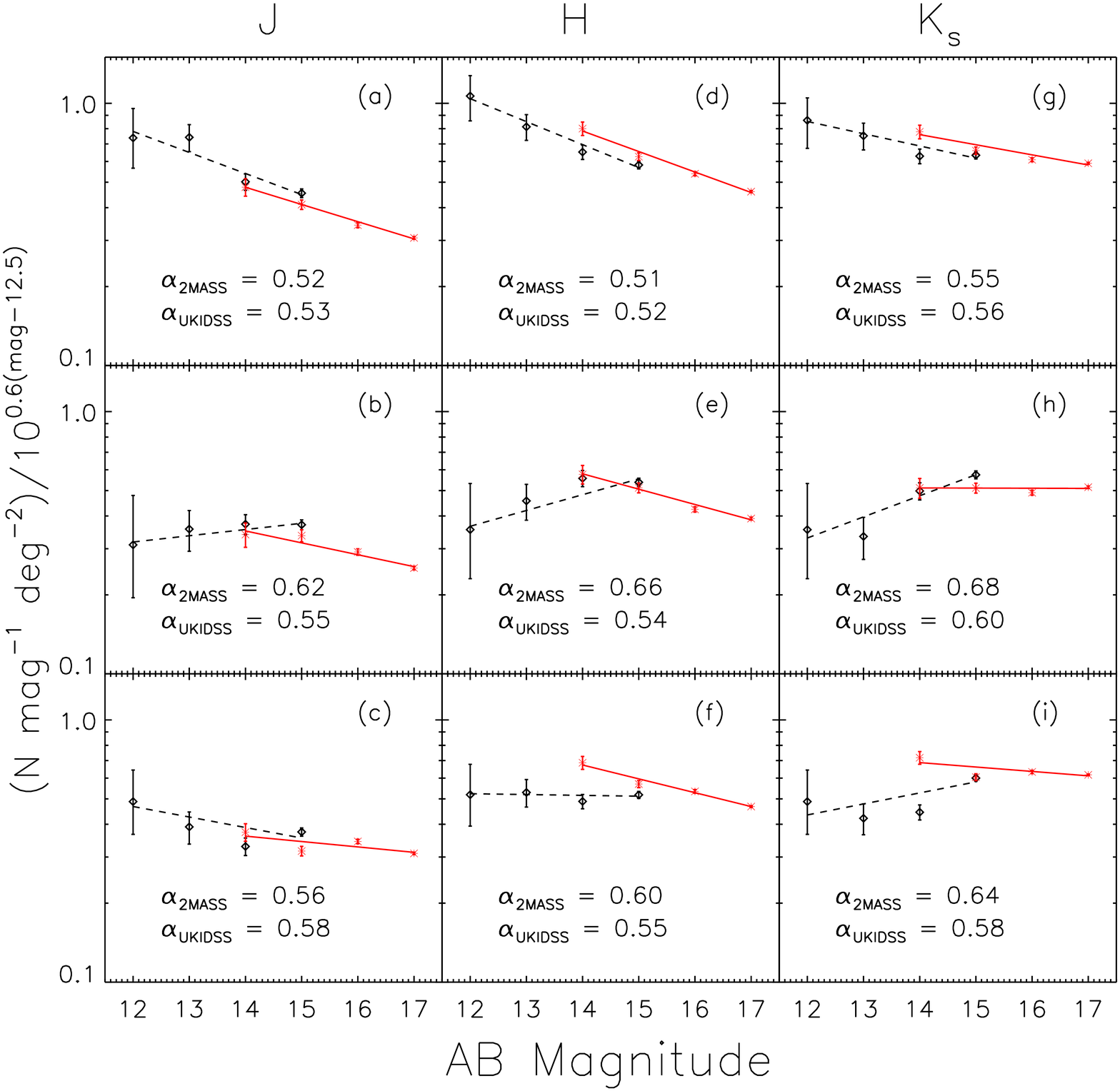}
\caption{\label{massvukidss} The slope of the 2MASS counts (black diamonds) in the
  range $11.5<JHK_s<15.5$ compared with the slope of the UKIDSS counts (red
  asterisks) in the range  $13.5<JHK_s<17.5$.  (a-c) show $J-$band counts, (d-f)
  show $H-$band counts, and (g-i) show $K_s-$band.  The top row of plots is for
  subfield 1, the middle for subfield 2, and the bottom for subfield 3 (see
  Table~\ref{ukidsstab} for subfield details).  The dashed and solid lines
  show an error-weighted least-squares fit to the 2MASS and UKIDSS counts,
  respectively.  The slopes for each are given on the plots and fitted errors on
  the slopes are $\sim 0.01$ in all cases.  The result is that for
  three large fields totalling $\sim 135$ deg$^2$ at high supergalactic
  latitude, no abrupt steepening of the counts slope is seen.}
\end{center}
\end{figure*}

We now consider the implications of our bright galaxy counts in terms of local
large scale structure.  The most striking feature in the local large scale structure is the
supergalactic Plane discovered by \citealt{deva53}.  The plane is
roughly perpendicular to our Galactic plane.  In Figure~\ref{masscomp}, we 
find that in the CLANS and CLASXS fields, the slope appears
distinctly sub-Euclidean, whereas in the other three fields the slopes are
closer to the Euclidean prediction.   This is  interesting because the CLANS
and CLASXS fields are of nearly
equatorial supergalactic latitude ($-1.8$ and $-3.9$ deg, respectively; see
Table~\ref{summarytab}).  In fact, all of our fields are at relatively low
absolute supergalactic latitude (average $|\rm{SGB}|\sim 10$ deg), with A370 being the furthest
off the plane at $-25.7$ deg.

The fields of \citet{Huan97}, where a strong super-Euclidean slope was found
in the $K_s-$band, span absolute supergalactic latitudes from
$10<\rm{SGB}<54$ with an average of $\rm{SGB} \sim 30$ deg. As a verification
of the \citet{Huan97} result, we did another
comparison in which we extracted
2MASS galaxies from patches of 25 deg$^2$ centered on the \citet{Huan97}
fields.  First, however, we extracted 2MASS counts from the actual areas of the
\citet{Huan97} fields and found a zeropoint offset of $-0.3$ to their counts.  We show this in Figure~\ref{huang_corr}.  We applied
this offset to their counts before proceeding with the comparison to the
larger (25 deg$^2$) fields.  

We plot the zeropoint-adjusted counts from \citet{Huan97} with the average of
2MASS counts over the larger surrounding areas in Figure~\ref{huang_v_mass}. We
calculate a slope of $\alpha = 0.67$ for the combined dataset.  This value is
slightly smaller than that found using just the \citet{Huan97} datapoints
($\alpha = 0.69$), but still distinctly super-Euclidean.  Over the magnitude
range sampled, this would still represent roughly a factor of two increase in
the space density of galaxies as found by \citet{Huan97}.     

\citet{Gard93} found a super-Euclidean bright counts slope of $\alpha = 0.67$
in their survey of a collection of fields totalling $\sim 10$ deg$^2$
at an average absolute SGB of $\sim 25$.  \citet{Vais00} found a sub-Euclidean slope at
bright magnitudes in the 
$K_s-$band at high supergalactic latitude ($\rm{SGB}=40$), but this can likely
be explained by the fact that they had
three Abell clusters in their fields (A2168 at $z=0.06$; A2211 at $z=0.15$; and
A2213 at $z=0.15$), which likely dominate their galaxy counts at bright
magnitudes.  The effect of just one galaxy cluster in a similar sized field
can be observed in our data in Figure~\ref{mycounts}, where the A370 ($z=0.37$) counts show a large excess in the range $17<JHK_s<19$.  The three lower redshift
clusters in the \citet{Vais00} fields would produce large excesses at brighter
magnitudes, and indeed they find their counts to be in excess over all other
published studies at bright magnitudes.  

Thus, from the evidence above,  it seems that a large fraction of the
variability in the slope and absolute numbers between different bright counts
studies can be attributed to nearby large scale structure, be it individual
galaxy clusters or local superstructure.  When individual nearby galaxy
clusters are avoided, there appears to be evidence for the slope of bright
galaxy counts to be related to supergalactic latitude.

To investigate this further, we calculated the bright counts slope in 2MASS as a
function of position on the sky by dividing the $|\rm{b}|>30$ catalog into
slices in supergalactic latitude ranging in area from $\sim 500-2500$ deg$^2$ each.  As a result, we obtained slope
measurements in $J$, $H$, and $K_s$ for $-60<\rm{SGB}<60$. We plot the
average 2MASS slope as a function of SGB in Figure~\ref{supgalslope}.  In this
figure the Euclidean slope prediction has been subtracted from the measured
values such that the dashed horizontal line represents the expectation for a
Euclidean slope in all three bands.  $J$-band is represented in blue,
$H-$band in black, and $K_s-$band in red.  The error in the slope measurements
is $\sim 0.01$ in all cases.  We note that the similarity in slope
measurements between bands suggests that our slope determinations are robust.   

In Figure~\ref{supgalslope} the only distinctly sub-Euclidean
slopes are found near the supergalactic equator, consistent with results from
our CLANS and CLASXS fields, with all other latitudes
appearing to be super-Euclidean.   At high northern supergalactic latitudes we
find that slopes exceeding the Euclidean prediction by $>0.05$, such as those
found by \citet{Huan97} and \citet{Gard93}, are typical.  One interpretation of
these results would be that near the supergalactic plane, the very brightest counts are
populated with an excess of sources causing the
slope to drop, and above the plane a relative paucity of nearby sources above the plane
causing the slope to steepen.  The southern supergalactic latitudes do not
display the same sharp rise to super-Euclidean as in the north.  The magnitude
range sampled in these counts is $10.5<JHK_s<15.5$, which means that the
brightest magnitude bin is centered on typical galaxies at a distance of $\sim
50$ Mpc.  This means that toward high northern SGB, where slopes are $\sim
0.1$ above the Euclidean prediction, the space density of
galaxies at a distance of $50$ Mpc could be low by roughly a factor of two
compared to the space density $250-350$ Mpc distant.  In the southern
supergalactic cap the slopes are not as steep ($\sim 0.03$ above Euclidean),
but still represent a possible underdensity of $\sim 25\%$ at $50$ Mpc
relative to $250-350$ Mpc.  

Next we took a broad average of 2MASS counts in and out of the supergalactic
plane. In Table~\ref{supgalbias} we give the measured average slope for galaxy
counts in the supergalactic plane ($|\rm{SGB}|< 10$) and out of the plane.  We
find that on average the counts in the plane are sub-Euclidean, consistent
with what we found in Figure~\ref{supgalslope}.  The slope out of the plane is
steeper by  some $\Delta\alpha \sim 0.02-0.06$ in both the northern and southern
supergalactic caps, and super-Euclidean on average. Thus, the expected
variability in the 2MASS bright counts slope is $\sim \pm 0.04$ due only to the very largest local structure.  

We next divided the 2MASS $|\rm{b}|>30$ sky into 844
patches of $\sim 25$ deg$^2$ each. We measured the slope of the galaxy counts from
$11<JHK_s<15$ for each and found an rms variability in the slope of $\sim 0.1$.  Inside
distances of $\sim 250$ Mpc, 2MASS is $1-2$ magnitudes deeper than $M^*$ and, as
such, is measuring a fairly complete sample of the local luminosity function.
This implies our 25 deg$^2$ patches are sampling structure on size scales
of $\sim 20$ Mpc and find rather strong variability.

\begin{deluxetable}{lccc}

\tabletypesize{\scriptsize}
\tablewidth{0pt}

\tablecaption{\label{supgalbias} 2MASS Galaxy Counts Slope Over the Galactic
  Caps, the Supergalactic Equator and the Supergalactic Caps} 

\tablehead{\textbf{Region} &\textbf{$J$ Slope}\tablenotemark{a} &\textbf{$H$ Slope}\tablenotemark{a} & \textbf{$K_s$ slope}\tablenotemark{a}} 

\startdata

Gal. $|\rm{b}|>30$~deg & 0.59 & 0.59 & 0.63 \\
$|\rm{SGB}|<10$~deg  & 0.55 & 0.57 & 0.59 \\
SGB$<-10$~deg  & 0.60 & 0.59 & 0.63 \\
SGB$>10$~deg  & 0.61 & 0.61 & 0.65 \\
$|\rm{SGB}|>10$~deg  & 0.61 & 0.60 & 0.64\\

\enddata
\tablenotetext{a}{Errors in fitted slopes are $\sim 0.01$ in all cases}

\end{deluxetable}

To test for the angular size over which all local large scale apart from the supergalactic
plane could be sufficiently averaged out, we divided up the 2MASS
$|\rm{b}|>30$ sky into increasingly larger subfields and measured the rms
variability in the measured slope across all subfields.  Figure~\ref{areaverr}
shows the result of this exercise for subfields ranging from $\sim 25-10,000$
deg$^2$.  We find that when averaging over subfields larger than $\sim 1000$
deg$^2$, the rms variability between all subfields is of the same order of
that induced by the supergalactic plane alone, and as such, all other local
large scale structure has been sufficiently averaged out.  In the 2MASS
sample, this suggests that effects of local large scale structure have been
sufficiently averaged out over scales of $\sim 150$ Mpc.  

Based on the analysis above, we conclude that the local supergalactic plane
is overdense by $10-15$\% compared to regions at distances of $\sim 250-350$
Mpc, while the voids just above and below the plane are underdense by up to a
factor of two relative to similarly distant regions.

Cold Dark Matter simulations such as the Millenium Run \citep{Spri05} predict
that the largest structures in the universe should be $\leq 100$ Mpc in
extent.  Observed large scale structures, most notably the
Sloan Great Wall \citep{Gott05}, demonstrate the existence of structure on
scales much larger than simulations predict.  Consensus has not yet been reached
on whether a structure like the Sloan Great Wall represents an extreme
non-linearity in the matter distribution of the universe, or if
inhomogeneities on several hundred Mpc scales are typical.  Two recent
analyses of the distribution of matter in the SDSS DR6 dataset
\citep{Sark09,Kita09} demonstrate the current range of interpretation of
observations, in which \citet{Sark09} arrive at scale of homogeneity of $70$
Mpc and \citet{Kita09} claim to have discovered a new large void $150$ Mpc in
diameter.  Clearly the typical size of large scale structure still depends on
how one interprets the data.  In our analysis, we find that averaging galaxy
counts over areas on the sky spanning $\sim 150$ Mpc sufficiently smooths out
local structure.  However, in the same data, the super-Euclidean slopes above
and below the supergalactic plane suggest that the local space density of
galaxies may be low by $25-100$\% compared to regions a few hundred Mpc away,
which would suggest that local structure exists on much larger scales than
$150$ Mpc.

\subsection{Comparison with the UKIDSS Large Area Survey}

We employ the UKIDSS survey to investigate the possibility that the entire 2MASS volume resides in a relatively
underdense region that is $\geq 250$ Mpc in radius, as suggested by
\citet{Huan97}.  The expectation if this were the case would be a steepening
of the galaxy counts slope in the range $15<JHK_s<17$, as, in fact, is seen in
the \citet{Huan97} sample when compared with 2MASS.  

As a test for a large void, we employ the UKIDSS \citep{Lawr07}
third data release (DR3; Warren et al., in preparation) Large Area Survey (LAS), which has recently become
public.  The DR3 LAS covers several hundred degrees of sky within the SDSS
footprint to depths of $JHK_s\sim 19$.  We selected  three large subfields totalling $\sim 135$ deg$^2$ in
the UKIDSS LAS.  We downloaded catalogs from the WFCam Science Archive, and the
range in celestial coordinates and SGB for these subfields are given in
Table~\ref{ukidsstab}.  We selected the fields to span a wide range of SGB above
and below the supergalactic plane.

We cross-correlated the UKIDSS catalogs with their SDSS counterparts and removed
stars from the catalogs via the SDSS classifier and $g^{\prime}$$JK$ color-color diagrams
in a similar manner to what is described in Section~\ref{sgsep}.  A comparison
between the slope of the 2MASS counts and the UKIDSS counts for the three
subfields is shown in Figure~\ref{massvukidss}.  2MASS counts are shown in
black diamonds and UKIDSS counts in red asterisks.  The dashed and solid lines
show error-weighted least-squares fits to the 2MASS and UKIDSS counts,
respectively.  The slopes for each fit are listed in the plots, and the
associated errors in the fitted slopes are $\sim \pm 0.01$ in all cases.  

The result is that for these three subfields, we do not observe a steepening
of the slope at magnitudes beyond the 2MASS limits.  In fact, in the majority
of cases in Figure~\ref{massvukidss}, the slope appears to be decreasing into
the fainter magnitudes of the UKIDSS survey. 
\begin{deluxetable}{lccc}[!ht]

\tabletypesize{\tiny}
\tablewidth{0pt}

\tablecaption{\label{ukidsstab} Three Subfields Selected from the UKIDSS DR3 LAS} 

\tablehead{\textbf{Subfield} &\textbf{$\Delta$RA (deg)}
  &\textbf{$\Delta$DEC (deg)}& \textbf{$\Delta$SGB (deg)}}

\startdata

1 &$201.5-222.1$& $9.0-11.4$ & $10-30$ \\
2 &$337.5-360.0$ & $-1.0-1.0$ & $13-34$ \\
3 &$135.0-150.0$ & $7.0-11.4$ & $-(37-52)$ \\

\enddata

\end{deluxetable}

\section{Summary}

\label{summary}

We have presented a deep, wide-field NIR survey over five widely separated fields
at high galactic latitude covering a total  of $\sim 3$ deg$^2$ in $J$, $H$,
and $K_s$. The deepest parts of the survey ($\sim 0.25$ square degrees) reach
5 $\sigma$ limits  $JHK_s > 24 $. As
such, this is one of the deepest wide-field NIR imaging surveys to date.  In
this paper we focus on the bright galaxy counts slope and the implications for
local large scale structure.  We leave analysis of the faint galaxy counts and
other applications of these data for future papers.  

We measure the slope of the bright galaxy counts in our data combined with
the larger 2MASS fields at our positions on the sky, and we find it consistent with the
Euclidean prediction, on average, over our five fields.  We note that our fields
are at a low average supergalactic latitude (SGB $\sim 10$) and our fields
near the supergalactic equator show a sub-Euclidean slope. In the 2MASS
$|\rm{b}|>30$ sky, we find the slope of the counts to be sub-Euclidean near
the supergalactic equator and super-Euclidean at all other SGB.  This is consistent both with our measured slope near the
supergalactic equator and with other studies that have found steeper slopes
above the plane, except when nearby galaxy clusters are included in the
counts.  This can be understood, at least in part, in terms of the fact that
our galaxy counts beginning at 11$^{th}$ magnitude are sampling the structure of the plane
itself in sightlines along the plane, and sampling the voids above and below
the plane in sightlines away from the plane.  

We further explore local large scale structure in the 2MASS sample by varying
the area over which the counts are averaged to find the angular scale on which
the variability between one area and another is of the same order as that
expected to be induced by the supergalactic plane alone.  We find that fields
of $\geq 1000$ deg$^2$ achieve this result, corresponding to averaging over
scales of $\sim 150$ Mpc in the 2MASS volume.  This is to say, we find
that on scales of $\sim 150$ Mpc, local large scale structure is
sufficiently averaged out of the counts.   However, our result that the galaxy
counts slope is super-Euclidean above and below the supergalactic plane
implies that the local space density of galaxies (away from the plane itself)
could be low by $25-100$\% relative to regions a few hundred Mpc distant.
This suggests that local structure may exist on scales much larger than $150$
Mpc.  

Finally, we explore the possibility of whether the entire 2MASS volume exists
in a region of relative underdensity through a comparison with the UKIDSS DR3
LAS.  We find that the slope of the UKIDSS counts is either consistent with
2MASS or takes a lower value over the three large subfields we selected.  This
result is inconsistent with the expectation of a steepening of the counts
slope in the range $15<JHK_s<17$ that would be expected if we lived inside a
large void of radius $\sim 300$ Mpc.  

Above all else, the comparison and analysis of several large NIR surveys
presented here has served to highlight the complexity of deciphering local
large scale structure from galaxy counts alone.  The completion of surveys
like UKIDSS will expand our view of the local Universe, but as demonstrated
here, even surveys of large areas on the sky may suffer biases due to local
structure.  As such, mapping out local large scale structure remains a complex
problem that will likely remain ambiguous until extensive spectroscopy in
combination with large scale photometric surveys can generate a three
dimensional picture of the local Universe.

\acknowledgements{We gratefully acknowledge support from NSF grants
AST 0239425 and AST 0708793 (A.~J.~B.) and AST 0407374 and AST 0709356
(L.~L.~C.), the University of Wisconsin Research Committee with funds granted
by the Wisconsin Alumni Research Foundation, and the David and Lucile Packard
Foundation (A.~J.~B.).  L.~T. was supported by a National Science Foundation
Graduate Research Fellowship and a Wisconsin Space Grant Consortium Graduate
Fellowship during portions of this work.  R.~C.~K. was supported by a
Wisconsin Space Grant Consortium Graduate Fellowship during portions of this
work. 

The authors thank the anonymous referee for their careful review of this paper
and insightful comments and suggestions that helped to improve the paper.  

This publication makes use of data products from the
Two Micron All Sky Survey (2MASS), which is a joint project of the University
of Massachusetts and the Infrared Processing and Analysis Center/California
Institute of Technology, funded by the National Aeronautics and Space
Administration and the National Science Foundation.  This publication also
makes use of data products from the Ukirt Infrared Deep Sky Survey (UKIDSS)
and the Sloan Digital Sky Survey (SDSS).

Funding for the SDSS and SDSS-II has been provided by the Alfred P. Sloan Foundation, the Participating Institutions, the National Science Foundation, the U.S. Department of Energy, the National Aeronautics and Space Administration, the Japanese Monbukagakusho, the Max Planck Society, and the Higher Education Funding Council for England. The SDSS Web Site is http://www.sdss.org/.

The SDSS is managed by the Astrophysical Research Consortium for the Participating Institutions. The Participating Institutions are the American Museum of Natural History, Astrophysical Institute Potsdam, University of Basel, University of Cambridge, Case Western Reserve University, University of Chicago, Drexel University, Fermilab, the Institute for Advanced Study, the Japan Participation Group, Johns Hopkins University, the Joint Institute for Nuclear Astrophysics, the Kavli Institute for Particle Astrophysics and Cosmology, the Korean Scientist Group, the Chinese Academy of Sciences (LAMOST), Los Alamos National Laboratory, the Max-Planck-Institute for Astronomy (MPIA), the Max-Planck-Institute for Astrophysics (MPA), New Mexico State University, Ohio State University, University of Pittsburgh, University of Portsmouth, Princeton University, the United States Naval Observatory, and the University of Washington.}

\bibliography{my}

\end{document}